\documentclass[aps,superscriptaddress,twocolumn,amsmath,amsfonts,amssymb]{revtex4-2} 
\usepackage{graphicx}
\usepackage{dcolumn}
\usepackage{bm}
\usepackage{xcolor}
\usepackage[colorlinks=true,linkcolor=blue,citecolor=blue,urlcolor=blue]{hyperref}
\usepackage{mathcomp}
\usepackage{here}
\usepackage{float}

\begin{document}

%\preprint{APS/123-QED}
\title{Exotic vortex states and enhancement of orbital upper critical fields in an $s$-wave quasicrystalline superconductor}

\author{Takumi Fukushima}  
\email{fukushima.17@osu.edu}
\affiliation{Department of Physics, The Ohio State University, Columbus, OH 43210, USA}
\affiliation{Institute for Solid State Physics, University of Tokyo, Kashiwa, Chiba 277-8581, Japan} 
\author{Yuki Yamamoto}
\email{yamamoto.yuki.5p3@ecs.osaka-u.ac.jp}
\affiliation{\mbox{Department of Physics, Graduate School of Science, The University of Osaka, Toyonaka, Osaka 560-0043, Japan}}
\author{Nayuta Takemori}
\email{nayuta.takemori.sci@osaka-u.ac.jp}
\affiliation{\mbox{Department of Physics, Graduate School of Science, The University of Osaka, Toyonaka, Osaka 560-0043, Japan}}
\affiliation{\mbox{Center for Quantum Information and Quantum Biology, The University of Osaka, Toyonaka, Osaka 560-0043, Japan}}
\affiliation{\mbox{Forefront Research Center, The University of Osaka, Toyonaka, Osaka 560–0043, Japan}}
\author{Shiro Sakai}
\affiliation{Physics Division, Sophia University, Chiyoda-ku, Tokyo 102-8554, Japan}
\author{Masanori Ichioka}
\affiliation{Research Institute for Interdisciplinary Science, Okayama University, Okayama, Okayama 700-8530, Japan}
\author{Anuradha Jagannathan}
\affiliation{Laboratoire de Physique des Solides, CNRS, Universit\'{e} Paris-Saclay, 91405 Orsay, France}

\date{\today}
            
%%%%%%%%%%%%%%%%%%%%%%%%%%%%%%%%%%%%%%%%%%%%%
\begin{abstract}

Quasicrystals possess long-range order without translational symmetry, giving rise to intrinsically inhomogeneous superconducting states.
Using the Bogoliubov-de Gennes mean-field theory, we investigate vortex states in an $s$-wave quasicrystalline superconductor on the Ammann-Beenker tiling, with particular emphasis on the high-field regime. 
We find that the vortex configuration can differ qualitatively from the Abrikosov triangular lattice. In particular, at filling $\nu=0.8$, vortices preferentially penetrate weak-pairing regions centered on sites with coordination number $8$, indicating that their spatial organization is strongly influenced by the local geometry.
Furthermore, we observe an enhanced upper critical field. 
At $\nu=0.8$, where the enhancement is particularly pronounced, the spatially averaged order parameter exhibits a plateau-like field dependence accompanied by the formation of vortex clusters.
These clusters accommodate additional flux in locally weak superconducting regions, providing a microscopic interpretation of the plateau-like behavior.
Moreover, the upper critical field increases upon cooling toward zero temperature without a clear sign of saturation.

\end{abstract}

\maketitle

%%%%%%%%%%%%%%%%%%%%%%%%%%%%%%%%%%%%%%%%%%%%%%%%%%%%%%%%%%%%%%%%%%%
\section{Introduction}
\label{sec:intro}
 
Quasicrystals are characterized by long-range structural order without translational symmetry and can exhibit rotational symmetries forbidden in periodic crystals \cite{Shechtman1984, Levine1984}.
This unique geometry gives rise to electronic states distinct from both the extended Bloch waves in periodic systems \cite{Bloch1928, Ashcroft1976} and the localized states characteristic of disordered systems \cite{Anderson1958, Lee1985}.
Specifically, the energy spectrum in quasicrystals is often singularly continuous, and the corresponding wave functions exhibit critical behavior \cite{Kohmoto1983, Ostlund1983, Niu1986, Tsunetsugu1986, Kohmoto1987, Tokihiro1988, Tsunetsugu1991}.
These electronic properties are expected to significantly influence the macroscopic quantum phases that emerge in quasicrystalline systems.
 
Superconductivity in quasicrystals has attracted considerable experimental and theoretical interest.
Bulk superconductivity was
first reported in the Bergman-type Al-Zn-Mg quasicrystal with $T_{\mathrm c}\simeq0.05$~K \cite{Kamiya2018}, and more recently in the van der Waals layered $\mathrm{Ta}_{1.6}\mathrm{Te}$ quasicrystal with $T_{\mathrm c}\simeq1$~K \cite{Tokumoto2024, Terashima2024, Matsudaira2026}.
In parallel, theoretical investigations have been conducted using real-space formulations, as the absence of translational symmetry precludes the conventional Bardeen-Cooper-Schrieffer (BCS) description in momentum space \cite{Bardeen1957, deGennes1999, Tinkham2004, Kita2015}.
The calculations based on the attractive Hubbard model have revealed that the superconducting order parameter is intrinsically spatially inhomogeneous, strongly reflecting the underlying quasiperiodic geometry \cite{Sakai2017, Araujo2019, Takemori2020, Liu2022, Fukushima2023proceeding, Fukushima2023, Liu2023, Junmo2025}.

While the zero-field superconducting state has been extensively studied, the behavior of quasicrystalline superconductors under magnetic fields remains comparatively unexplored, despite some pioneering theoretical works \cite{Sakai2019, Nagai2022, Tomayeva2026}.
In these systems, the formation of the Abrikosov triangular lattice \cite{Abrikosov1957, Kleiner1964, Essmann1967, Hess1989} is challenged by the intrinsic quasiperiodic geometry.
Although previous studies have addressed the intrinsic pinning of a single vortex \cite{Nagai2022} or spatially sign-alternating order parameters \cite{Sakai2019}, (i) how vortices collectively organize under strong magnetic fields and (ii) how this organization affects the orbital upper critical field $B_{{\mathrm c}2}$ remain unclear. The intrinsic spatial inhomogeneity of the order parameter suggests that vortices preferentially enter weakly superconducting regions to reduce condensation-energy loss. This can qualitatively reshape the vortex configuration and potentially modify $B_{{\mathrm c}2}$. Clarifying these issues is timely in light of recent experimental progress on magnetic field responses in quasicrystalline superconductors \cite{Tokumoto2024, Terashima2024, Matsudaira2026}.

To address these issues, we investigate vortex states in an $s$-wave superconductor on the Ammann-Beenker (AB) tiling under magnetic field, which is a paradigmatic two-dimensional quasiperiodic structure with eightfold rotational symmetry. By self-consistently solving the Bogoliubov-de Gennes (BdG) equations, we analyze the superconducting order parameter in both real and perpendicular spaces, with particular emphasis on the high-field regime. 
For comparison, we also perform the corresponding calculations on a square lattice (SL) under comparable conditions.
We find that the vortex configuration can differ qualitatively from the conventional Abrikosov triangular lattice. At filling $\nu=0.8$, vortices preferentially penetrate weak-pairing regions centered on sites with coordination number 8, indicating that their spatial organization is strongly influenced by the local geometry.
We further find that the orbital upper critical
field $B_{{\mathrm c}2}$ is enhanced relative to that of the SL. At $\nu=0.8$, where this enhancement is particularly pronounced, the spatially averaged order parameter exhibits a plateau-like field dependence, accompanied by the formation of vortex clusters.
Such clustering allows additional flux to be accommodated in locally weak superconducting regions, providing a microscopic interpretation of the plateau-like behavior.
Furthermore, $B_{{\mathrm c}2}$ increases upon cooling toward zero temperature without a clear sign of saturation.
 
The rest of this paper is organized as follows.
In Sec.~\ref{sec:model}, we introduce the attractive Hubbard model on the AB tiling and explain the BdG formalism.
We present our results: (i) the vortex states with a single flux quantum in Sec.~\ref{subsec:SC states}, (ii) the vortex states in high-field regimes in Sec.~\ref{subsec:vortex}, (iii) the enhancement of $B_{{\mathrm c}2}$ in Sec.~\ref{subsec:Bc2}, and (iv) the temperature dependence of $B_{{\mathrm c}2}$ in Sec.~\ref{subsec:T-Bc2}.
Finally, Sec.~\ref{sec:summary} summarizes our findings.

%%%%%%%%%%%%%%%%%%%%%%%%%%%%%%%%%%%%%%%%%%%%%%%%%%%%%%%%%%%%%%%%%%%%%%
\section{Model and Method}
\label{sec:model} 

%%%%%%%%%%%%%%%%%%%%%%%%%%%%%%%%%%%%%%%%%%%%%%%%%%%%%%%%%%%%%%%%%%%%%%
\subsection{Ammann-Beenker tiling and its square approximant}
\label{subsec:AB tiling}

\begin{figure}[t]
\centering
\includegraphics[width=\linewidth]{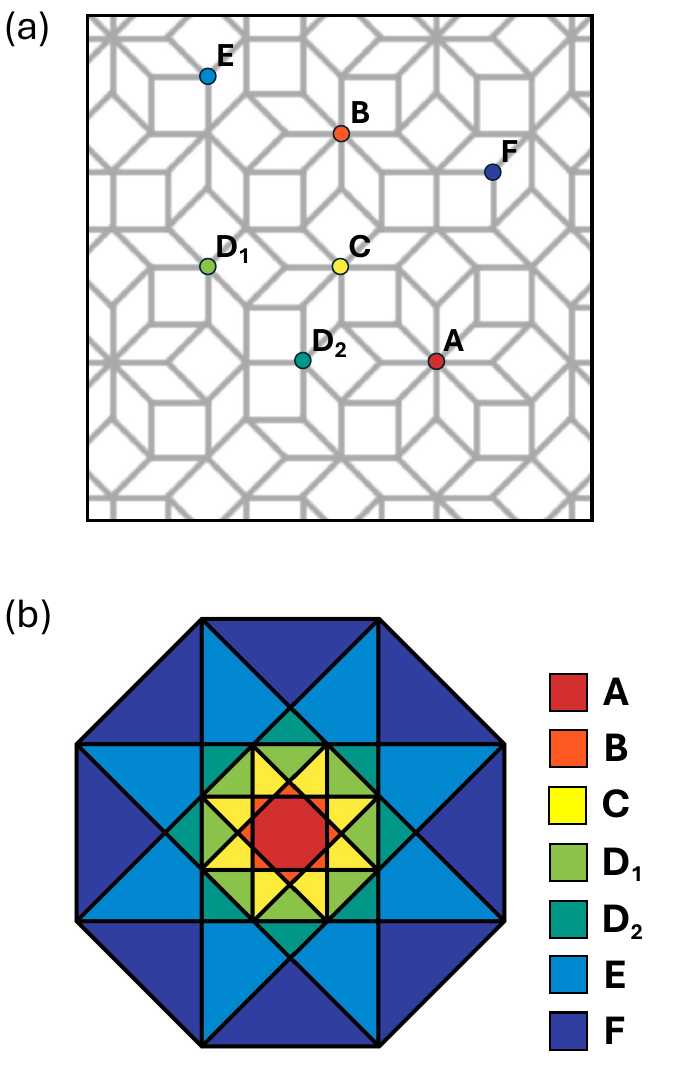}
\caption{(a) Portion of the AB tiling showing representative sites of the seven vertex types. (b) Perpendicular space colored based on the vertex type projected onto the corresponding position. The correspondence between the vertex types and colors is shown in panel (a).}
\label{fig:ABtiling}
\end{figure}

We introduce the Ammann-Beenker tiling [Fig.~\ref{fig:ABtiling}(a)], which is a prototypical two-dimensional quasiperiodic tiling \cite{Beenker1982,Ammann1992}. The coordination number $Z_i$ at each vertex ranges from $3$ to $8$. The vertices are grouped into six coordination classes, labeled $\text{A}$ through $\text{F}$ for $Z_{i}=8, 7,\cdots,3$, respectively. The $Z_i=5$ class (type $\text{D}$) is further divided into $\text{D}_1$ and $\text{D}_2$ according to the neighboring vertex types, giving seven vertex types in total \cite{Jagannathan2004,Jagannathan2005}.

The AB tiling is generated by the cut-and-project method, in which a four-dimensional hypercubic lattice is projected onto a two-dimensional subspace \cite{Duneau1989,Jagannathan2024}.
Each lattice point \(\bm{R}\in\mathbb{Z}^4\) is written as
\begin{equation}
\label{eq:vecR}
\bm{R}=\sum_{\mu=1}^{4} n_\mu \bm{a}_\mu,
\qquad
n_\mu \in \mathbb{Z},
\end{equation}
where \(\bm{a}_\mu\) are the fundamental translation vectors. The four-dimensional space is decomposed into two orthogonal subspaces: the physical space \(S_{\parallel}\), spanned by \(\bm{e}_1\) and \(\bm{e}_2\), and the perpendicular space \(S_{\perp}\), spanned by \(\bm{e}_3\) and \(\bm{e}_4\). These bases are related by
\begin{equation}
\label{eq:relationae} 
\begin{pmatrix}
\bm{a}_1 & \bm{a}_2 & \bm{a}_3 & \bm{a}_4
\end{pmatrix}
=
\begin{pmatrix}
\bm{e}_1 & \bm{e}_2 & \bm{e}_3 & \bm{e}_4
\end{pmatrix}
C,
\end{equation}
where the transformation matrix $C$ is defined as
\begin{equation}
\label{eq:matrixC} 
C=\frac{1}{\sqrt{2}}
\begin{pmatrix}
\sqrt{2} & 1 & 0 & -1 \\
0 & 1 & \sqrt{2} & 1 \\
\sqrt{2} & -1 & 0 & 1 \\
0 & 1 & -\sqrt{2} & 1
\end{pmatrix}.
\end{equation}
The selected four-dimensional lattice points are projected onto \(S_{\parallel}\) to form the vertices of the AB tiling. The corresponding projections onto \(S_{\perp}\) reflect the local geometries of the vertices, such that vertices with similar geometries in \(S_{\parallel}\) are mapped to nearby regions in \(S_{\perp}\). Consequently, \(S_{\perp}\) is partitioned into domains associated with distinct vertex types, as shown in Fig.~\ref{fig:ABtiling}(b). This correspondence makes the perpendicular-space representation useful for analyzing the dependence of physical properties on local geometry.

For numerical calculations, we use a finite square approximant to represent the infinite quasiperiodic AB tiling \cite{Jagannathan2024}. It is constructed using the cut-and-project method with the transformation matrix \(C_n\), given by
\begin{equation}
\label{eq:matrixCn}
C_n=\frac{P_n}{\sqrt{2}N_n} \begin{pmatrix} 
\sqrt{2}\alpha_n & \sqrt{2} & 0 & -\sqrt{2} \\ 
0 & \sqrt{2} & \sqrt{2}\alpha_n & \sqrt{2} \\
2 & -\alpha_n & 0 & \alpha_n \\
0 & \alpha_n & -2 & \alpha_n
\end{pmatrix}.
\end{equation}
Here, $\alpha_n = 1 + P_{n-1}/P_n$, where $P_n$ is the $n$-th Pell number, and $N_n$ is given by $2N_n^2 = 4P_n^2 + (-1)^n$.
With the present normalization, \(C/\sqrt{2}\) and \(C_n/\sqrt{2}\) are orthogonal matrices. This normalization sets the tile-edge length of the infinite quasiperiodic AB tiling to unity, while the corresponding edge lengths in the square approximant are close to unity. In the limit of $n \to \infty$, we obtain the original transformation matrix $C$ as 
\begin{equation}
\label{eq:limitCn}
\lim_{n\to\infty}C_n=C.
\end{equation}
In the present study, we employ the $n=4$ square approximant. This system consists of $N=1393$ sites with a side length $L=2N_4\approx33.97$.
For comparison with a periodic crystalline system, we also perform the same calculations on an $L=30$ square lattice containing $N=900$ sites \cite{Wang1995,Takigawa2004}.

%%%%%%%%%%%%%%%%%%%%%%%%%%%%%%%%%%%%%%%%%%%%%%%%%%%%%%%%%%%%%%%%%%%%%%
\subsection{The model Hamiltonian}
\label{subsec:model}

Assuming $s$-wave superconducting states, we consider the attractive Hubbard model on the AB tiling in the presence of a uniform magnetic field ${\bm B}=B{\hat z}$, applied perpendicular to the system. The Hamiltonian reads
\begin{align}
\label{eq:hamiltonian}  
      \hat{\cal H} = 
      -\sum_{\langle i,j \rangle, \sigma} &
      \left(t_{i,j} {\rm e}^{-{\rm i}\theta_{i,j}} \hat{c}^\dag_{i\sigma} \hat{c}_{j\sigma} + \rm{H.c.} \right) \notag \\ 
       & + U \sum_{i} \hat{n}_{i\uparrow} \hat{n}_{i\downarrow} - \mu \sum_{i,\sigma} \hat{n}_{i\sigma},
\end{align}
where $U<0$ is the onsite attraction.
The fermion operator $\hat{c}^{(\dag)}_{i\sigma}$ annihilates (creates) an electron with spin $\sigma$ at site $i$. We assume hopping only between sites connected by the edges of the rhombic or square tiles, denoted as $\langle i,j \rangle$, with a uniform amplitude $t_{i,j} = t$, and we set $t$ as the unit of energy. 
In the non-interacting case, the energy width of the density of states is approximately $8.5t$ \cite{Sakai2021}. 
We define the local electron density $n_i=\sum_\sigma \langle \hat{n}_{i\sigma} \rangle$ with $\hat{n}_{i\sigma}=\hat{c}_{i\sigma}^\dag \hat{c}_{i\sigma}$ and tune the chemical potential $\mu$ to fix the filling $\nu=(1/N)\sum_{i}n_{i}$. % with the site number $N$.
Here $\langle \cdot \rangle$ denotes the expectation value at temperature $T$.
In this study, we fix the interaction strength to $U=-2$ and select two fillings, $\nu=0.5$ and $0.8$. 

We incorporate a magnetic field through the Peierls substitution for the hopping amplitude $t_{i,j}$ \cite{Peierls1997}, in which the Peierls phase is 
\begin{equation}
\label{eq:Peierls}
    \theta_{i,j}=\frac{e}{\hbar}\int_{\bm r_j}^{\bm r_i}\bm A(\bm r)\cdot \mathrm{d}\bm r .
\end{equation}
We choose the Landau gauge, $\bm A(\bm r)=(0,Bx,0)$, and set $e=\hbar=1$ hereafter. 
%Furthermore, we impose periodic boundary conditions and treat the square approximant of linear size $L$ as a magnetic unit cell, which fixes the field strength in the following way \cite{Wang1995,Takigawa2004}:
%\begin{equation}
%    B=\frac{\Phi_0}{L^2}N_\phi,
%\end{equation}
%so that $B$ is controlled by the integer $N_\phi\in\mathbb{N}$. In our units, the flux quantum and the superconducting flux quantum are $\Phi_0=2\pi$ and $\Phi^{\rm SC}_0=\pi$, respectively.
We impose periodic boundary conditions and treat the entire system as a square magnetic unit cell. 
%For comparison with a periodic crystalline system, we also perform the same calculations on an $L=30$ square lattice containing $N=900$ sites \cite{Wang1995,Takigawa2004}.
The primitive vectors of the unit cell are
\begin{equation}
    \boldsymbol{u}_x = (L,0),
    \qquad
    \boldsymbol{u}_y = (0,L).
\end{equation}
Consistency of the periodic boundary conditions requires the total flux through the unit cell to be an integer multiple of the electronic flux quantum $\Phi_0=2\pi$.
Accordingly, the magnetic field is restricted to
\begin{equation}
    B = \frac{\Phi_0}{L^2} N_\phi,
    \qquad
    N_\phi \in \mathbb N.
\label{eq:magnetic_field}
\end{equation}
In our units, the superconducting flux quantum is $\Phi_0^{\rm SC}=\pi$, and hence the unit cell contains
$2N_\phi$ vortices.

%The fermion operator $\hat{c}^{(\dag)}_{i\sigma}$ annihilates (creates) an electron with spin $\sigma$ at site $i$. We assume hopping only between %nearest-neighbor sites, 
%\tf{sites connected by the edges of the rhombic or square tiles}, denoted as $\langle i,j \rangle$, with a uniform amplitude $t_{i,j} = t$, and we set $t$ as the unit of energy. In the non-interacting case, the energy width of the density of states is approximately $8.5t$ \cite{Sakai2021}. 
%We define the local electron density $n_i=\sum_\sigma \langle \hat{n}_{i\sigma} \rangle$ with $\hat{n}_{i\sigma}=\hat{c}_{i\sigma}^\dag \hat{c}_{i\sigma}$ and tune the chemical potential $\mu$ to fix the filling $\nu=(1/N)\sum_{i}n_{i}$. % with the site number $N$.
%\tf{Here $\langle \cdot \rangle$ denotes the expectation value at temperature $T$.}
%In this study, we fix the interaction strength to $U=-2$ and select two fillings, $\nu=0.5$ and $0.8$. 

Using the BdG mean-field theory \cite{deGennes1999, Zhu2016}, one can reduce the Hamiltonian to 
\begin{equation}
\begin{gathered}
    \label{eq:Hmf-1}
    \hat{\cal H} =
    \sum_{i,j}
    \left(\begin{array}{cc}
    \hat{c}^\dagger_{i\uparrow} & 
    \hat{c}_{i\downarrow}\end{array}\right)
    {\cal H}_{i,j}
    \left(\begin{array}{c}
    \hat{c}_{j\uparrow}\\
    \hat{c}^{\dag}_{j\downarrow}\end{array}\right), \\ 
%    \label{eq:Hmf-2}
    {\cal H}_{i,j} =
    \left( \begin{array}{cc} 
    K_{\uparrow i,j }  & \Delta_i \delta_{i,j} \\
    \Delta^\ast_i \delta_{i,j} &  - K^\ast_{ \downarrow i,j} 
\end{array} \right),
\end{gathered}
\end{equation} 
where $K_{\sigma i,j}=-t{\rm e}^{-{\rm i}\theta_{i,j}}+(U n_{i\bar\sigma}-\mu)\delta_{i,j}$ is the kinetic term with $n_{i \sigma}=\langle \hat{n}_{i \sigma} \rangle$ and $\Delta_i=-U\langle {\hat c}_{i\uparrow}{\hat c}_{i\downarrow} \rangle$ is a site-dependent superconducting order parameter.  
Under a translation by a lattice vector $\boldsymbol{R}=m\boldsymbol{u}_x+n\boldsymbol{u}_y$ ($m,n \in \mathbb Z$), the order parameter satisfies
\begin{equation}
    \Delta(\boldsymbol{r}_i+\boldsymbol{R})
    =
    e^{-2i\boldsymbol{A}(\boldsymbol{R})
    \cdot\boldsymbol{r}_i}
    \Delta(\boldsymbol{r}_i).
\label{eq:magnetic_boundary_condition}
\end{equation}
%The Hartree term needs to be explicitly incorporated because it exhibits site dependence. Therefore, it cannot be absorbed into the chemical potential term, as done in Ref.~\cite{Sakai2017, Takemori2020, Fukushima2023}.
The BdG Hamiltonian in Eq.~\eqref{eq:Hmf-1} is diagonalized through the Bogoliubov transformation 
\begin{align}
\left(\begin{array}{c}
   {\hat c}_{i\uparrow}\\
   {\hat c}^{\dag}_{i\downarrow} \\
   \end{array} \right)
   =\sum_{\epsilon}
   \left(\begin{array}{cc}
       u_{\epsilon}({\bm r}_{i}) & -v^{*}_{\epsilon}({\bm r}_{i}) \\
        v_{\epsilon}({\bm r}_{i})&u^{*}_{\epsilon}({\bm r}_{i})  \\
   \end{array}\right)
   \left( \begin{array}{c} 
     {\hat \gamma}_{\epsilon \uparrow} \\ 
     {\hat \gamma}^{\dag}_{\epsilon \downarrow}\\
   \end{array} \right).     
\end{align}
Finally, we obtain the BdG equation,
\begin{align}
\label{eq:BdG}
    \sum_{j} {\cal H}_{i,j} 
    \left( \begin{array}{c}
    u_{\epsilon}({\bm r}_j) \\ 
    v_{\epsilon}({\bm r}_j)\\
    \end{array} \right) 
    = E_\epsilon 
    \left( \begin{array}{c}
    u_{\epsilon}({\bm r}_i) \\ 
    v_{\epsilon}({\bm r}_i) \\ 
   \end{array} \right).
\end{align}
In Eq.~\eqref{eq:BdG}, $E_{\epsilon}$ denotes an eigenenergy of the BdG Hamiltonian ${\cal H}_{i,j}$, and $u_{\epsilon}(\bm{r}_{i})$ and $v_{\epsilon}(\bm{r}_{i})$ denote wave functions on site $i$. The index $\epsilon$ distinguishes eigenstates of ${\cal H}_{i,j}$, which runs from 1 to $2N$, including eigenstates with both positive and negative $E_{\epsilon}$. 
%As the self-consistent condition, the gap equation and the local electron density for each spin state are 
The self-consistency conditions for the gap and the local electron density of each spin state are
\begin{align}
\label{eq:gap-eq}
    \Delta_i &= - U \sum_{\epsilon} u_{\epsilon}({\bm r}_{i}) v^{\ast}_{\epsilon}({\bm r}_{i})(1-f(E_{\epsilon})), \\
\label{eq:density1-eq}
    n_{i \uparrow} &= \sum_\epsilon |u_{\epsilon}({\bm r}_{i})|^2 f(E_{\epsilon}), \\
\label{eq:density2-eq}
    n_{i \downarrow} &= \sum_\epsilon |v_{\epsilon}({\bm r}_{i})|^2 (1-f(E_{\epsilon})),
\end{align}
The Fermi-Dirac distribution function at temperature $T$ is $f(E)=1/(e^{E/T}+1)$, where we set $k_{\mathrm B}=1$.

The self-consistent calculation is initialized with a configuration containing $2N_\phi$ vortices that are compatible with the periodic boundary conditions. This configuration is used only as an initial condition; the converged solution is not constrained to retain the initial vortex arrangement and can reorganize according to the lattice geometry and vortex-vortex interactions.
Since we assume the Type-II limit, the magnetic field inside the superconductor is treated as a constant value equal to the external magnetic field.
Hence, the Meissner state and a finite lower critical field $B_{{\mathrm c}1}$ are not captured; effectively $B_{{\mathrm c}1}=0$, and any nonzero applied field induces a vortex state.

%%%%%%%%%%%%%%%%%%%%%%%%%%%%%%%%%%%%%%%%%%%%%%%%%%%%%%%%%
\section{Results}
\label{sec:result}

%%%%%%%%%%%%%%%%%%%%%%%%%%%%%%%%%%%%%%%%%%%%%%%%%%%%%%%%%
\subsection{Vortex states at $N_\phi=1$} 
\label{subsec:SC states}

The zero-field $s$-wave superconducting states on quasiperiodic tilings have been investigated in several theoretical studies \cite{Sakai2017, Araujo2019, Takemori2020, Liu2022, Fukushima2023proceeding, Fukushima2023, Liu2023}. 
These studies revealed that the order parameter $\Delta_i$ varies spatially even among sites with the same coordination number $Z_i$, due to variations in second- and further-neighbor environments.
Building on these results, we first examine the vortex state at $N_\phi=1$, corresponding to one electronic flux quantum, as previously studied in Ref.~\cite{Nagai2022}. 

\begin{figure}[h]
    \centering
    \includegraphics[width=\linewidth]{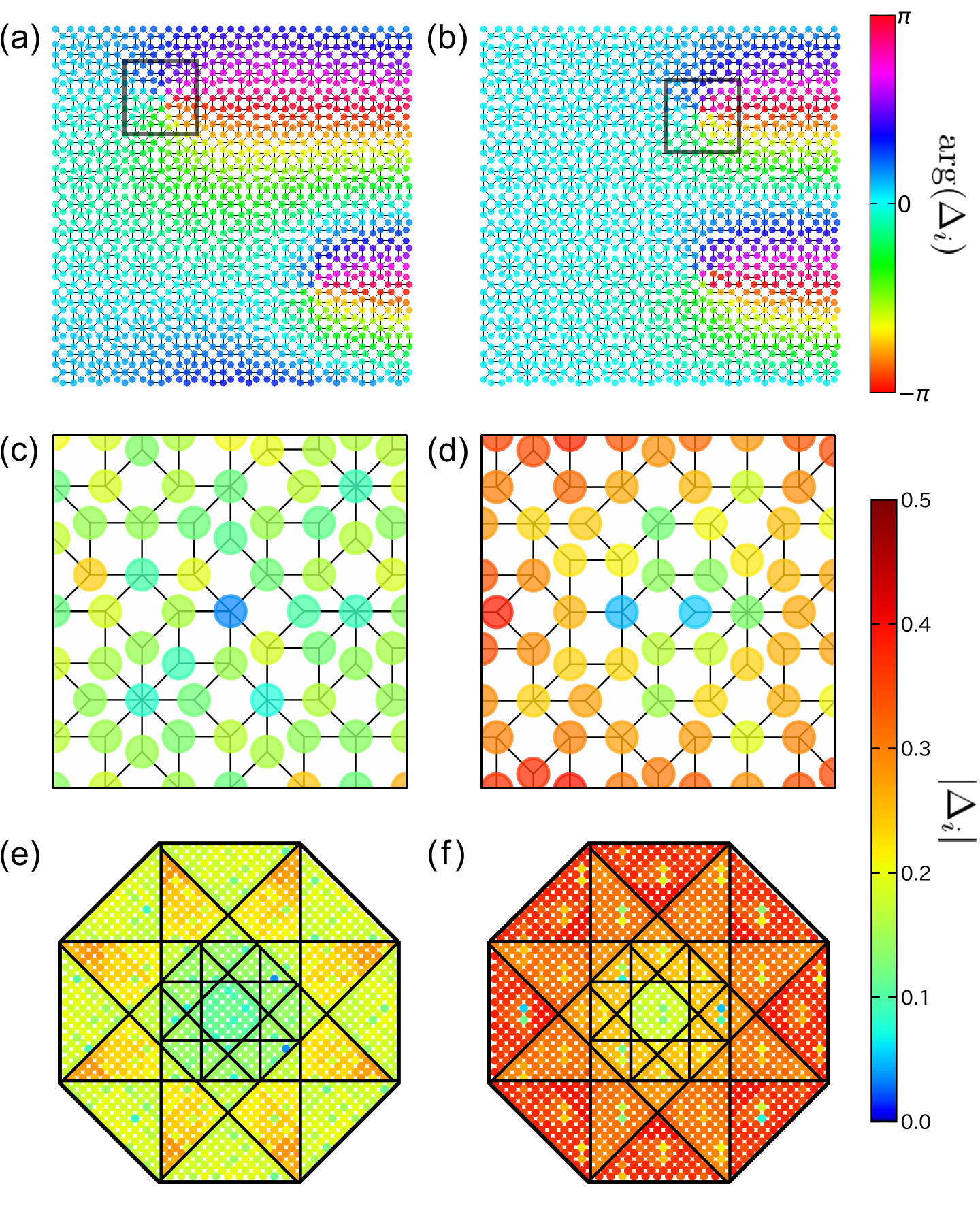}
    \caption{Spatial distributions of the order parameter $\Delta_i$ in real space [panels (a)-(d)] and perpendicular space [panels (e) and (f)] at $T=0.01$ for $\nu=0.5$ [panels (a), (c), and (e)] and $\nu=0.8$ [panels (b), (d), and (f)] at $N_\phi=1$.
    In the top panels, the color of the circles represents the phase $\arg(\Delta_i)$, and the black boxes indicate the regions enlarged in the middle panels. The middle and bottom panels show the amplitude $|\Delta_i|$.
    % In the top panels, the size and the color of the circles represent the amplitude $|\Delta_i|$ and the phase $\arg(\Delta_i)$, respectively. The bottom panels show $|\Delta_i|$.
    }
    \label{fig:OP_B=2}
\end{figure}

Figure \ref{fig:OP_B=2} shows the phase of $\Delta_i$ in real space [panels (a) and (b)], enlarged real-space distributions of its amplitude [panels (c) and (d)], and the amplitude in perpendicular space [panels (e) and (f)].
% Figure \ref{fig:OP_B=2} shows the spatial distributions of $\Delta_i$ in real space [panels (a) and (b)] and perpendicular space [panels (c) and (d)].
The results were obtained by self-consistently solving Eqs.~(\ref{eq:BdG})-(\ref{eq:density2-eq}) at $T=0.01$ for two fillings $\nu=0.5$ [panels (a), (c) and (e)] and $0.8$ [panels (b), (d) and (f)]. 
% In the top panels, the size and color of the circles represent the amplitude $|\Delta_i|$ and the phase $\arg(\Delta_i)$, respectively.
A vortex is identified by a $2\pi$ phase winding, as seen within the black boxes in panels (a) and (b), where the color map completes one full cycle around the vortex core.
%A vortex is identified by a $2\pi$ phase winding, which appears as one complete cycle of the color map. 
%Two such phase-winding centers are observed because one electronic flux quantum, $\Phi_0=2\pi$, corresponds to two superconducting flux quanta, $\Phi_0^{\rm SC}=\pi$.
As mentioned in Sec.~\ref{subsec:model}, due to the periodic boundary conditions, two vortices appear in this case. The panels (c) and (d) show the vicinity of one of these vortices.

In the top and middle panels, we identify vortex cores as regions in which the phase winds by $2\pi$ while $|\Delta_i|$ is strongly suppressed. %The vortices appear at specific types of sites depending on the filling. 
When the vortex-core sites identified in real space are mapped onto perpendicular space, they fall within pockets where $|\Delta_i|$ is nearly zero, most prominently in the vertex-$\text{D}_1$ and vertex-$\text{F}$ sites.
In particular, the vortex penetration at vertex-$\text{D}_1$ sites is consistent with Ref.~\cite{Nagai2022}.
%Having identified the core locations, we next compare the sizes of the cores. For $\nu=0.5$, the region of suppressed $|\Delta_i|$ around each phase singularity extends over a relatively wide area, indicating a comparatively large vortex core. By contrast, for $\nu=0.8$, $|\Delta_i|$ recovers over a shorter distance from the vortex center, indicating a smaller vortex core. This qualitative trend is consistent with the BCS relation, $\xi \sim 1/\Delta$ \cite{Bardeen1957}.
Panel (e) additionally shows a pronounced variation of $|\Delta_i|$ within the vertex-$\text{E}$ region, where $|\Delta_i|$ increases toward the corners of the octagonal domain.
%This feature reflects the intrinsic zero-field inhomogeneity of the superconducting state, in which sites with $Z_i=4$ support relatively large values of $|\Delta_i|$ \cite{Fukushima2023}.
This variation indicates that the pairing amplitude depends on local environments beyond the coordination number alone. It has been shown that, for $\nu=0.5$, vertex-$\text{E}$ sites support relatively large values of $|\Delta_i|$ in the zero-field state \cite{Fukushima2023}.

\begin{figure*}[t]
    \centering
    \includegraphics[width=1.0\textwidth]{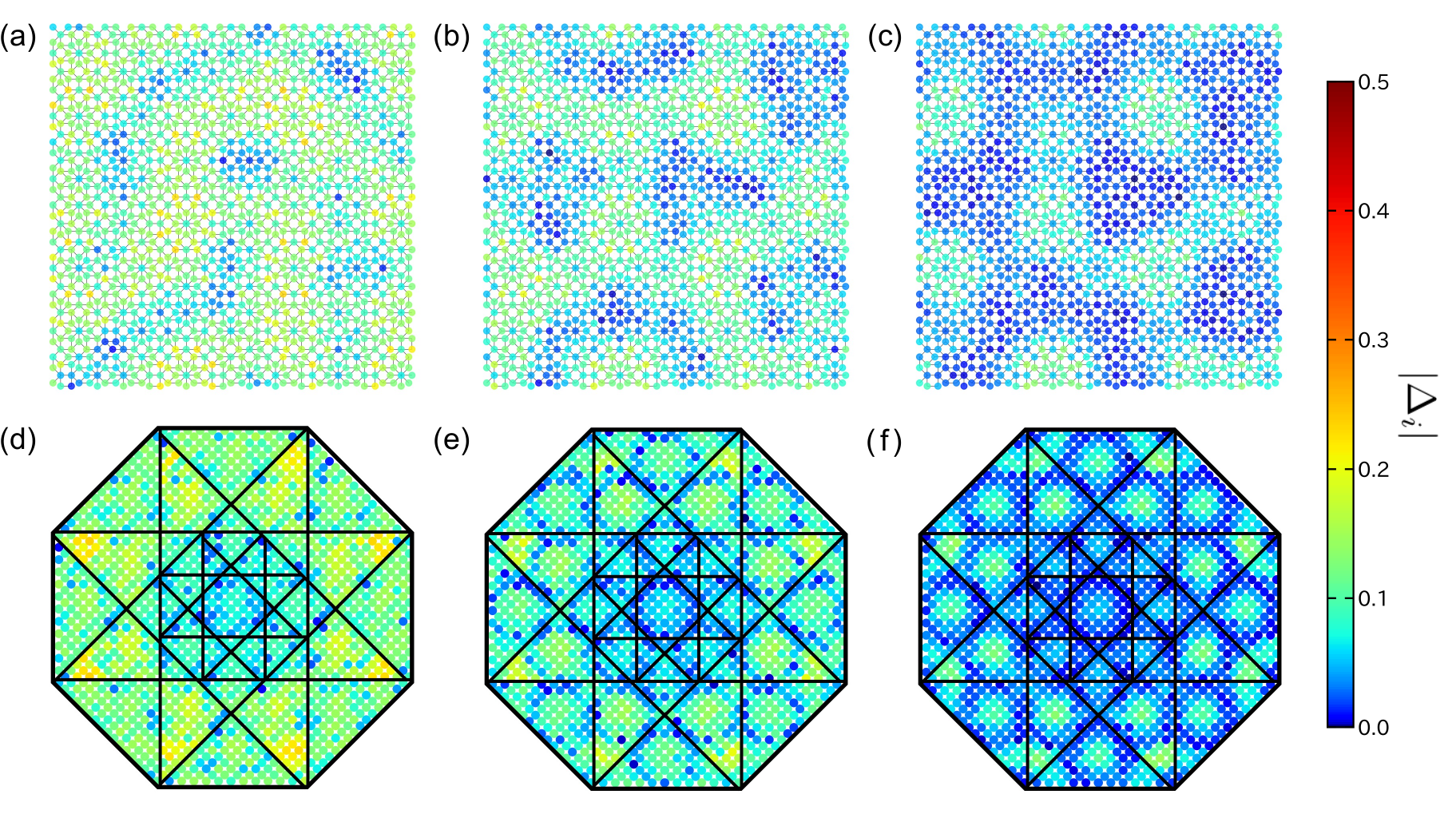}
    \caption{Spatial distributions of the order parameter amplitude $|\Delta_i|$ in real space [panels (a)-(c)] and perpendicular space [panels (d)-(f)] at $T=0.01$ for $\nu=0.5$. The left, middle, and right panels correspond to $B\approx0.5B_{\rm{c}2}$ ($N_\phi=10$), $0.7B_{\rm{c}2}$ ($N_\phi=15$), and $0.9B_{\rm{c}2}$ ($N_\phi=20$), respectively.}
    \label{fig:OP_T=0.01_n=0.5}
\end{figure*}

Due to the lack of translation symmetry, the coherence length can depend on the local environment in quasiperiodic systems. %cannot be determined uniquely in quasiperiodic systems. 
We therefore estimate an effective coherence length $\xi$ by fitting the spatial profile of the superconducting order-parameter amplitude to the following Ginzburg-Landau (GL) form \cite{Tinkham2004}:
\begin{align}
    \Delta(r) = \bar{\Delta} \tanh \left(\frac{r}{\xi}\right),
\label{eq:GL_coherence}
\end{align}
where $r$ is the distance from the vortex center, and $\bar{\Delta}$ is the spatially averaged order parameter defined by $\bar{\Delta}=(1/N)\sum_i |\Delta_i|$.
The value of $\bar{\Delta}$ and the resulting $\xi$ are summarized in Table \ref{tab:gap_coherence_length}.
The effective $\xi$ is larger at $\nu=0.5$ than at $\nu=0.8$, consistent with the broader vortex-core profile observed in Fig.~\ref{fig:OP_B=2} and with the qualitative BCS relation $\xi\sim1/\Delta$ \cite{Bardeen1957}.
%Using them, the fit yields $\xi_{\rm AB}=4.35$ for $\nu=0.5$ and $\xi_{\rm AB}=2.39$ for $\nu=0.8$.
%For comparison, in the $L=30$ square lattice (SL) under identical conditions, we extract $\xi_{\rm SL}=4.09$ for $\nu=0.5$, and $\xi_{\rm SL}=1.81$ for $\nu=0.8$. 
%Although these estimates are subject to finite-size effects and fitting uncertainty, the relative magnitude of $\xi_{\rm AB}$ and $\xi_{\rm SL}$ shows no systematic trend with filling. The fitted values satisfy $\xi_{\rm SL}<\xi_{\rm AB}$ at $\nu=0.5$ and $0.8$, whereas additional calculations at $\nu=0.6$ and $0.7$ yield the opposite ordering, $\xi_{\rm SL}>\xi_{\rm AB}$. Among the fillings examined, the absolute difference $|\xi_{\rm AB}-\xi_{\rm SL}|$ is largest at $\nu=0.8$.
We note that, sufficiently far from the vortex cores, the distribution of $|\Delta_i|$ remains essentially unchanged from its zero-field profile reported in Refs.~\cite{Araujo2019, Fukushima2023}. Intrinsic vortex pinning in this low-field case was previously reported in Ref.~\cite{Nagai2022}. We therefore focus on the high-field regime near the upper critical field $B_{{\mathrm c}2}$ in the following sections.

\begin{table}[h]
\caption{Spatially averaged superconducting order parameter $\bar{\Delta}$ and effective coherence length $\xi$ obtained by fitting the isolated-vortex profiles to Eq.~\eqref{eq:GL_coherence} for the square lattice (SL) and the Ammann-Beenker (AB) approximant at $T=0.01$ and $N_\phi=1$ for fillings $\nu=0.5$ and $0.8$.}
\label{tab:gap_coherence_length}
\begin{ruledtabular}
\renewcommand{\arraystretch}{1.3}
\begin{tabular}{cccc}
Lattice & $\nu$ & $\bar{\Delta}$ & $\xi$ \\
\hline 
SL & $0.5$ & $0.12$ & $4.09$ \\
AB & $0.5$ & $0.19$ & $4.35$ \\
SL & $0.8$ & $0.31$ & $1.81$ \\
AB & $0.8$ & $0.31$ & $2.39$
\end{tabular}
\renewcommand{\arraystretch}{1.0}
\end{ruledtabular}
\end{table}

\begin{figure*}[t]
    \centering
    \includegraphics[width=\textwidth]{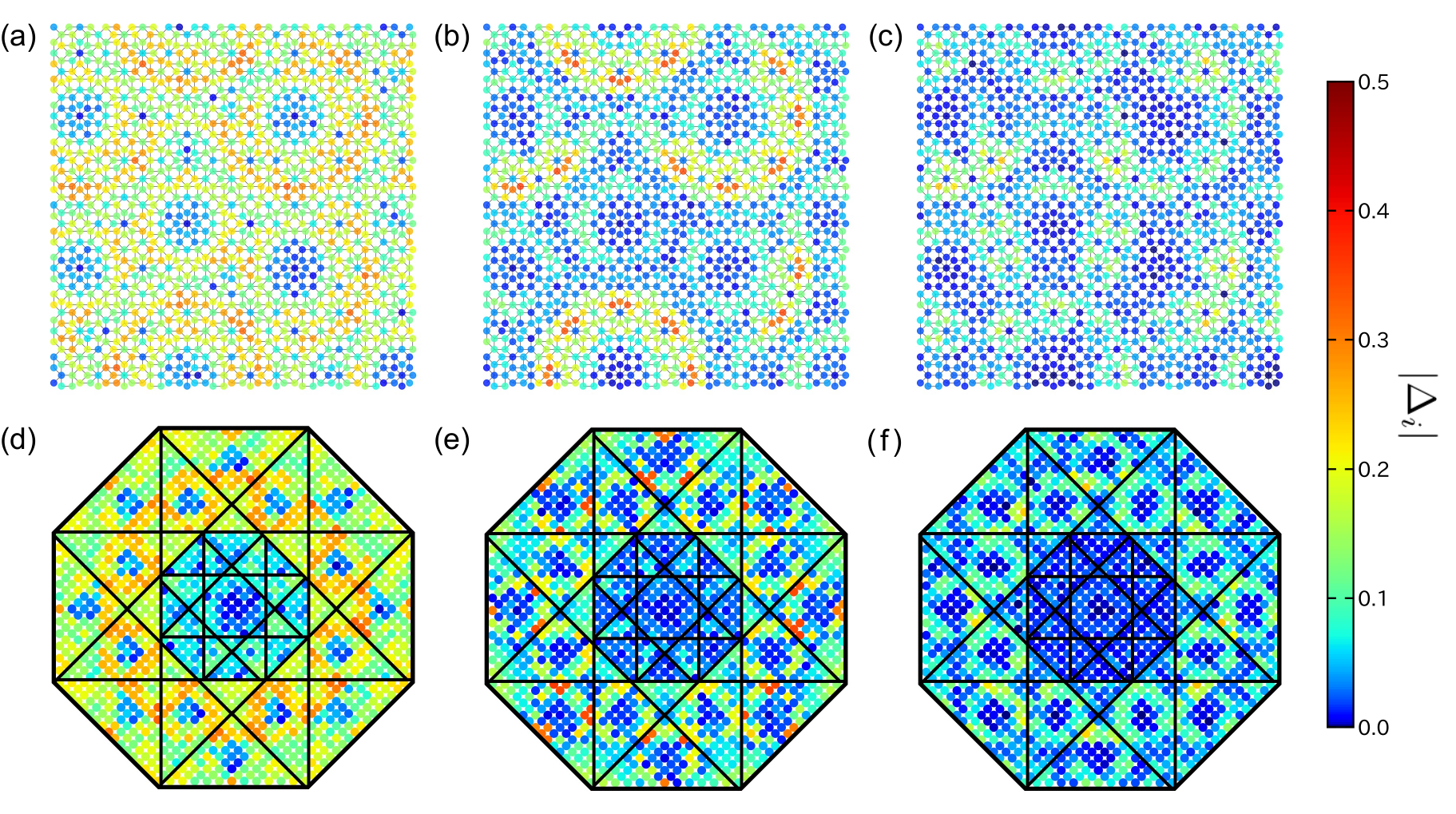}
    \caption{Spatial distributions of the order parameter amplitude $|\Delta_i|$ in real space [panels (a)-(c)] and perpendicular space [panels (d)-(f)] at $T=0.01$ for $\nu=0.8$. The left, middle, and right panels correspond to $B\approx0.5B_{\rm{c}2}$ ($N_\phi=65$), $0.7B_{\rm{c}2}$ ($N_\phi=90$), and $0.9B_{\rm{c}2}$ ($N_\phi=115$), respectively.}
    \label{fig:OP_T=0.01_n=0.8}
\end{figure*}

%%%%%%%%%%%%%%%%%%%%%%%%%%%%%%%%%%%%%%%%%%%%%%%%%%%%%%%%%
\subsection{Vortex states in the high-field regime}
\label{subsec:vortex}

We investigate the vortex states at a temperature of $T=0.01$ in the high-field regime defined by $0.5B_{\mathrm{c}2} \le B \le 0.9B_{\mathrm{c}2}$.
Figures \ref{fig:OP_T=0.01_n=0.5} and \ref{fig:OP_T=0.01_n=0.8} show the spatial distributions of the order parameter amplitude $|\Delta_i|$ in both real and perpendicular spaces for the fillings $\nu=0.5$ and $0.8$, respectively.
In each panel, regions where the amplitude is strongly suppressed ($|\Delta_i| \approx 0$) are clearly visible, indicating the locations of the vortices. As we see in the following, the vortex arrangement changes qualitatively with the filling: while the $\nu=0.5$ case remains relatively diffuse, for $\nu=0.8$ the vortices preferentially penetrate weak-pairing regions centered on vertex-A ($Z_i=8$) sites, and their spatial organization is strongly influenced by the local geometry.
%the $\nu=0.8$ case exhibits an octagonal vortex pattern reflecting the eightfold rotational symmetry of the underlying tiling.

For $\nu=0.5$, the strongly suppressed-$|\Delta_i|$ regions are broadly distributed in real space. In the zero-field limit, the vertex-E sites generally possess relatively large $|\Delta_i|$ \cite{Fukushima2023}, which provides a background inhomogeneity that also influences the high-field response. 
At $B\simeq0.5B_{\mathrm{c}2}$, superconductivity is locally suppressed in a diffuse and largely featureless manner in real space [Fig.~\ref{fig:OP_T=0.01_n=0.5}(a)]. %, indicating that the vortex \red{distribution remains spatially broad rather than exhibiting a clear preference for specific regions}.
%pattern remains spatially broad rather than \tf{preferring} a specific pattern. 
In perpendicular space [Fig.~\ref{fig:OP_T=0.01_n=0.5}(d)], large $|\Delta_i|$ remains mainly near each vertex of the octagon within the vertex-E region, and the characteristic feature found in the single-flux case of Fig.~\ref{fig:OP_B=2}(e) persists. 
At $B\simeq0.7B_{\mathrm{c}2}$, the suppressed-$|\Delta_i|$ regions expand and the overall superconductivity is further weakened in real space [Fig.~\ref{fig:OP_T=0.01_n=0.5}(b)]. %, while the suppression still appears spatially diffuse.
In perpendicular space [Fig.~\ref{fig:OP_T=0.01_n=0.5}(e)], superconductivity is suppressed particularly in the regions corresponding to vertices A through D$_2$, whereas the surviving superconducting domain in the vertex-E region splits into two parts; a similar splitting is also observed in the vertex-F region, where the domain divides into three parts. Near $B_{\mathrm{c}2}$, at $B\simeq0.9B_{\mathrm{c}2}$, superconductivity survives only in a few local clusters in real space [Fig.~\ref{fig:OP_T=0.01_n=0.5}(c)], which are typically around vertex-A-centered regions, and it is largely suppressed in other regions. In perpendicular space [Fig.~\ref{fig:OP_T=0.01_n=0.5}(f)], the separation of the superconducting regions in the vertex-E and F domains becomes even more evident, consistent with the strongly suppressed superconductivity seen in real space. Overall, while the suppression of $|\Delta_i|$ remains broadly distributed in real space at $\nu=0.5$, perpendicular space reveals a vertex-dependent pattern of superconducting suppression as $B$ approaches $B_{\mathrm{c}2}$. 

On the other hand, for $\nu=0.8$, vortices exhibit a clear preference for specific local environments. In particular, we find that $|\Delta_i|$ is strongly suppressed in regions centered on vertex-A sites.
As discussed in Refs.~\cite{Araujo2019, Fukushima2023}, $|\Delta_i|$ at vertex-A sites is inherently smaller even in the absence of a magnetic field, providing favorable regions for vortex penetration. 
At $B\simeq0.5B_{\mathrm{c}2}$, superconductivity is suppressed in clusters centered on vertex-A sites in real space [Fig.~\ref{fig:OP_T=0.01_n=0.8}(a)]. The remaining superconducting regions form connected ring-like structures. 
%In perpendicular space [Fig.~\ref{fig:OP_T=0.01_n=0.8}(d)], superconductivity is uniformly suppressed in the region spanning vertices A to D$_1$, whereas it survives in the region from D$_2$ to F. 
In perpendicular space [Fig.~\ref{fig:OP_T=0.01_n=0.8}(d)], the suppression of $|\Delta_i|$ is more extensive in the region spanning vertices A to D$_1$, while relatively large amplitudes remain over substantial portions of the D$_2$--F region, despite locally suppressed areas within the vertex-E and vertex-F regions. This spatial contrast reflects the zero-field dependence of $|\Delta_i|$ on the vertex types \cite{Fukushima2023}.
%Moreover, relatively stronger superconductivity appears near the boundaries of the inner octagons within these domains, forming a two-layer structure. Consequently, vortex regions in perpendicular space exhibit a nested structure located at the centers of octagonal superconducting regions. 
%At $B\simeq0.7B_{c2}$, the clusters with suppressed superconductivity grow in real space [Fig.~\ref{fig:OP_T=0.01_n=0.8}(b)], while the width of the ring structure decreases. Reflecting the eightfold symmetry of the tiling, superconductivity becomes relatively stronger at locations corresponding to the vertices of the octagon. 
At $B\simeq0.7B_{\mathrm{c}2}$, the suppressed clusters expand, and the superconducting regions become narrower [Fig.~\ref{fig:OP_T=0.01_n=0.8}(b)].
%In perpendicular space [Fig.~\ref{fig:OP_T=0.01_n=0.8}(e)], sites with particularly strong superconductivity are found in the vertex-F region, consistent with the enhanced superconductivity at the octagonal vertices in real space. 
%Correspondingly, relatively strong superconductivity survives mainly near the octagonal-vertex regions in real space and in the vertex-F region in perpendicular space [Fig.~\ref{fig:OP_T=0.01_n=0.8}(e)].
Relatively strong superconductivity remains at vertex-E and vertex-F sites near vertex-A sites in real space, with the same
tendency evident in perpendicular space [Fig.~\ref{fig:OP_T=0.01_n=0.8}(e)].
%Near the upper critical field, at $B\simeq0.9B_{c2}$, superconductivity persists only faintly near the octagonal-vertex regions in real space [Fig.~\ref{fig:OP_T=0.01_n=0.8}(c)] and is largely suppressed elsewhere. In perpendicular space [Fig.~\ref{fig:OP_T=0.01_n=0.8}(f)], superconductivity is completely suppressed in the octagonally symmetric region at the center, and only marginal superconductivity survives near its boundary, consistent with the feature already visible at $B\simeq0.5B_{c2}$ [Fig.~\ref{fig:OP_T=0.01_n=0.8}(d)]. 
%Near $B_{{\mathrm c}2}$, at $B\simeq0.9B_{{\mathrm c}2}$, superconductivity survives only faintly in these regions and is largely suppressed elsewhere [Figs.~\ref{fig:OP_T=0.01_n=0.8}(c) and \ref{fig:OP_T=0.01_n=0.8}(f)].
Near $B_{{\mathrm c}2}$, at $B\simeq0.9B_{{\mathrm c}2}$, superconductivity is strongly suppressed over most of the system in real space [Fig.~\ref{fig:OP_T=0.01_n=0.8}(c)]. In perpendicular space, relatively large $|\Delta_i|$ remains mainly in the vertex-E and vertex-F regions, with finite amplitudes also persisting in parts of the vertex-D$_2$ region [Fig.~\ref{fig:OP_T=0.01_n=0.8}(f)].
%Based on these observations for this filling, we conclude that the vortices preferentially penetrate weak-pairing regions centered on vertex-A ($Z_i=8$) sites, and that the resulting vortex distribution reflects the eightfold quasiperiodic organization of these local environments.
Overall, these results show that the high-field vortex organization is strongly influenced by the weak-pairing vertex-A-centered regions and the local quasiperiodic geometry.
The supercurrent distributions presented in Appendix \ref{app:supercurrent} further support the identification of these suppressed regions with vortex structures. Moreover, calculations under open boundary conditions in Appendix \ref{app:OBCs} show that the preferential vortex penetration into vertex-A-centered regions and the associated spatial organization persist even for a finite cluster without global eightfold symmetry. These results indicate that the observed vortex organization originates from the local geometry rather than from the boundary conditions or the global symmetry of the finite system.

%%%%%%%%%%%%%%%%%%%%%%%%%%%%%%%%%%%%%%%%%%%%%%%%%%%%%%%%%
\subsection{Upper critical field $B_{{\mathrm c}2}$ and vortex clustering}
\label{subsec:Bc2}

\begin{figure}[b]
    \centering
    \includegraphics[width=\linewidth]{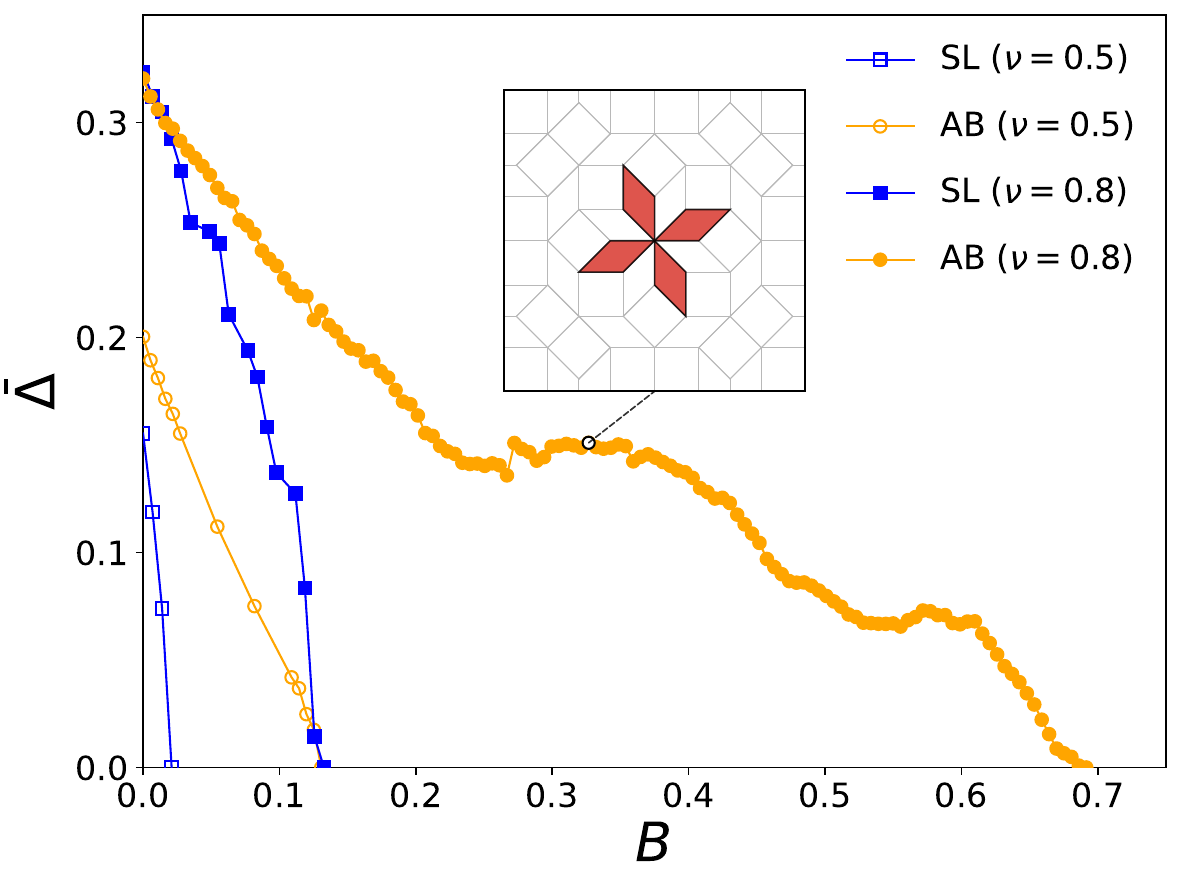}
    \caption{Field dependence of the spatially averaged order parameter $\bar\Delta$ for the square lattice (SL) and the Ammann–Beenker (AB) approximant at the fillings $\nu=0.5$ and $0.8$. The inset shows a representative vortex cluster in the plateau region of the AB approximant at $\nu=0.8$. The red plaquettes denote elementary plaquettes with winding number $W_p=1$.}
    \label{fig:Bc2_T=0.01}
\end{figure}

The upper critical field $B_{{\mathrm c}2}$ is identified as the magnetic field at which the spatially averaged order-parameter amplitude $\bar\Delta$ vanishes.
We compare $B_{{\mathrm c}2}$ for the AB approximant with that for the SL at the same filling.
Figure \ref{fig:Bc2_T=0.01} plots the field dependence of $\bar\Delta$ for both the AB approximant and the SL at $T=0.01$ for $\nu=0.5$ and $0.8$.
In the SL, $\bar\Delta$ is larger at $\nu=0.8$ than at $\nu=0.5$ due to the larger density of states at the Fermi level in the former, and the corresponding value of $B_{{\mathrm c}2}$ is also higher.
At both fillings, $\bar\Delta$ decreases smoothly with increasing magnetic field and approaches zero near $B_{{\mathrm c}2}$ within the field resolution of the present calculation. 
%Furthermore, ${\bar \Delta}$ exhibits mean-field behavior approaching zero at both fillings, indicating a second-order phase transition from the vortex state to the normal state.
For reference, we estimate $B_{{\mathrm c}2}$ from the effective coherence length $\xi$ obtained in Sec.~\ref{subsec:SC states} using the GL relation \cite{Tinkham2004},
\begin{align}
\label{eq:Bc2-xi}
    B_{{\mathrm c}2} = \frac{\Phi^{\mathrm{SC}}_0}{2\pi\xi^2}.
\end{align}
For the SL, this relation yields $B_{{\rm c}2}\sim0.03$ for $\nu=0.5$ and $B_{{\rm c}2}\sim0.15$ for $\nu=0.8$, in reasonable agreement with the numerical results shown in Fig.~\ref{fig:Bc2_T=0.01}. 

In contrast, the AB approximant exhibits a substantially larger $B_{{\mathrm c}2}$ than the SL at both fillings. 
The enhancement is particularly pronounced at $\nu=0.8$, where the low-field values of $\bar\Delta$ are nearly identical for the AB approximant and the SL.
Using the effective $\xi$ extracted from the isolated-vortex profiles in Sec.~\ref{subsec:SC states}, Eq.~\eqref{eq:Bc2-xi} gives $B_{{\mathrm c}2}\sim0.03$ for $\nu=0.5$ and $B_{{\mathrm c}2}\sim0.09$ for $\nu=0.8$. These values are considerably smaller than the numerically determined critical fields ($B_{{\mathrm c}2}\sim0.13$ for $\nu=0.5$ and $B_{{\mathrm c}2}\sim0.7$ for $\nu=0.8$) in Fig.~\ref{fig:Bc2_T=0.01}, particularly at $\nu=0.8$. This discrepancy indicates that the effective vortex-core size extracted in the low-field regime does not provide a direct measure of the length scale controlling $B_{{\mathrm c}2}$. 
In particular, the fitted $\xi$ characterizes the spatial recovery of the order parameter around an isolated vortex, whereas the instability near $B_{{\mathrm c}2}$ is governed by the response of the entire superconducting state to the magnetic field. These two length scales need not coincide in the intrinsically inhomogeneous AB approximant. Therefore, the low-field vortex-core size alone cannot account for the enhanced $B_{{\mathrm c}2}$.
Another notable feature at $\nu=0.8$ is the stepwise field dependence of $\bar\Delta$, in contrast to the smooth decrease observed in the SL. For $0<B\lesssim0.2$, $\bar\Delta$ decreases approximately linearly, followed by a plateau-like regime over $0.2\lesssim B\lesssim0.4$, where it remains nearly constant. It then resumes its decrease for $0.4\lesssim B\lesssim0.5$, with a smaller plateau appearing before superconductivity vanishes at $B_{{\mathrm c}2}$.

To identify the vortex positions, we evaluate the winding number on each elementary plaquette $p$,
\begin{equation}
    W_p = \frac{1}{2\pi}
    \left[\sum_{\langle i,j \rangle\in\partial p}
    \left(\phi_j-\phi_i - 2\theta_{i,j}\right) + 2\Phi_p \right],
    \label{eq:plaquette_winding}
\end{equation}
where $\phi_i=\arg(\Delta_i)$, and $\Phi_p=\sum_{\langle i,j\rangle\in\partial p}\theta_{i,j}$ is the magnetic flux through plaquette $p$. Here, each gauge-invariant phase difference in the sum is taken in the principal branch \((-\pi,\pi]\).
The tile-edge length is set to unity, so $\Phi_p=B$ for a square and $\Phi_p=B/\sqrt{2}$ for a rhombus. A plaquette with $W_p=1$ contains one elementary vortex.
Applying this analysis to the broadest plateau regime in Fig.~\ref{fig:Bc2_T=0.01}, we find that the vortices are not distributed independently throughout the system. As illustrated in the inset of Fig.~\ref{fig:Bc2_T=0.01}, some rhombic plaquettes with $W_p=1$ form a vortex cluster around a vertex-A site ($Z_i=8$). The separation between neighboring vortices in such a cluster is on the order of the edge length and is therefore much smaller than the effective coherence length $\xi_{\mathrm{AB}}\simeq 2.39$ estimated in Sec.~\ref{subsec:SC states}. Similar clusters are formed around many vertex-A sites throughout the system. 
Since the separation between neighboring vortices in a cluster is much smaller than the effective core size, the suppressed regions associated with neighboring vortices can substantially overlap. As further shown in Appendix \ref{app:plateau_suppressed_regions}, the spatial fraction of strongly suppressed superconducting regions changes only weakly over the plateau regime, even though the number of flux quanta continues to increase. Together with the winding-number analysis, this behavior supports the interpretation that additional vortices are preferentially accommodated within already weakened structural regions, allowing the surrounding robust superconducting regions to remain largely intact.
%Since the vortex cores within each cluster strongly overlap, the increase in the number of flux quanta does not lead to a proportional increase in the spatial extent of the regions where $\lvert\Delta_i\rvert$ is strongly suppressed. 
%Instead, additional vortices are accommodated primarily within local structural regions in which superconductivity is already weakened. Consequently, the total area of the strongly suppressed regions changes only weakly over the plateau regime, while the surrounding superconducting regions remain largely intact. 
This provides a microscopic interpretation of the plateau-like behavior of $\bar\Delta$ in Fig.~\ref{fig:Bc2_T=0.01}.

%%%%%%%%%%%%%%%%%%%%%%%%%%%%%%%%%%%%%%%%%%%%%%%%%%%%%%%%%
\subsection{Temperature dependence of upper critical field}
\label{subsec:T-Bc2}

\begin{figure}[t]
    \centering
    \includegraphics[width=\linewidth]{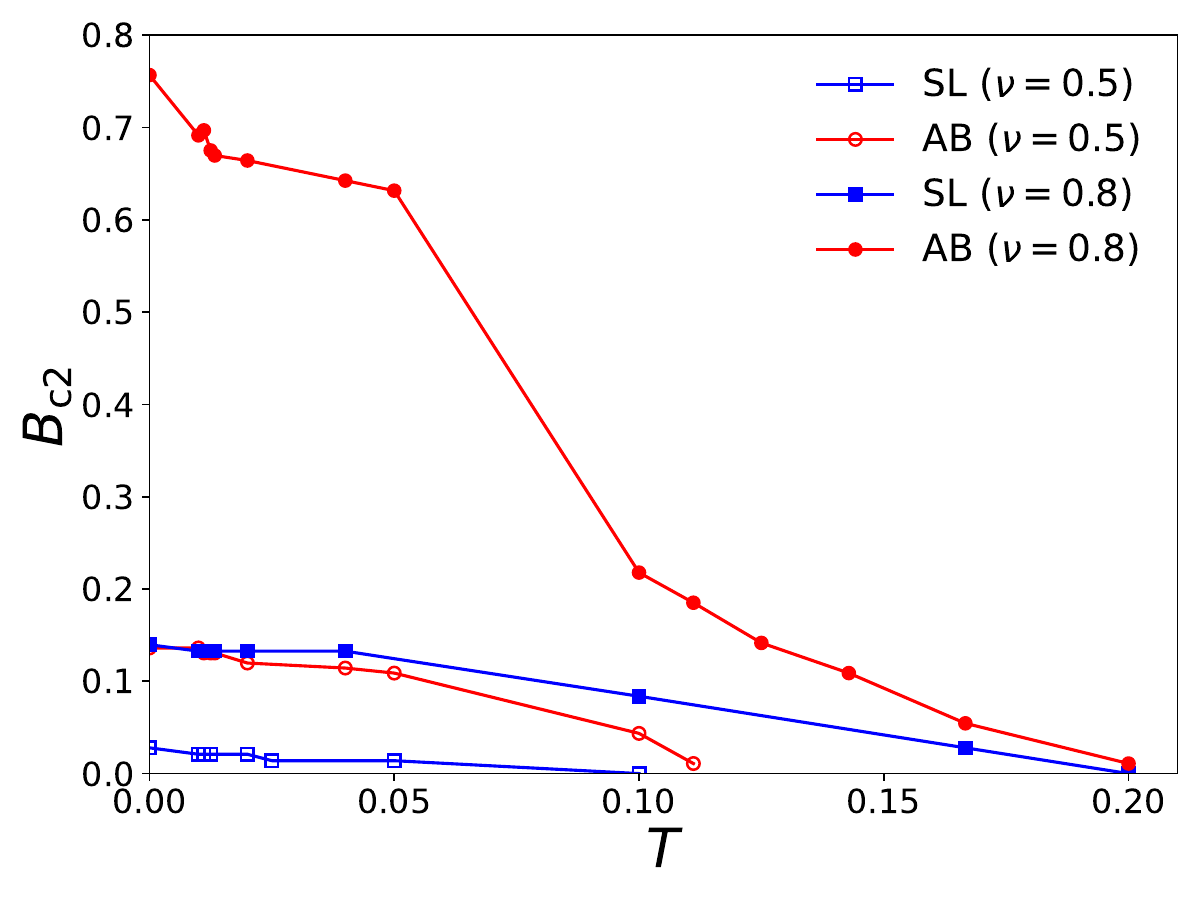}
    \caption{Temperature dependence of the upper critical field $B_{c2}(T)$ for the Ammann-Beenker (AB) approximant and the square lattice (SL) at fillings $\nu=0.5$ and $0.8$. Open and filled symbols denote $\nu=0.5$ and $0.8$, respectively.}
    \label{fig:Bc2-T}
\end{figure}

Figure \ref{fig:Bc2-T} compares the temperature dependence of $B_{{\mathrm c}2}(T)$ for the SL and the AB approximant at $\nu=0.5$ and $0.8$. At both fillings, the AB approximant exhibits a substantially larger $B_{{\mathrm c}2}(T)$ than the SL. 
The enhancement is particularly pronounced for $\nu=0.8$ near $T=0$, consistent with the low-temperature comparison in Fig.~\ref{fig:Bc2_T=0.01}. At this filling, $B_{{\mathrm c}2}(T)$ of the AB approximant continues to increase upon cooling toward zero temperature without any clear sign of saturation, whereas the corresponding variation in the SL is much weaker.
At $\nu=0.5$, the overall scale of $B_{{\mathrm c}2}(T)$ is substantially smaller than at $\nu=0.8$, but the AB approximant still exhibits an enhanced $B_{{\mathrm c}2}$ relative to the SL.

The filling dependence of $B_{{\mathrm c}2}(T)$ is consistent with the distinct high-field spatial responses discussed in Secs.~\ref{subsec:vortex} and \ref{subsec:Bc2}. At $\nu=0.5$, the suppression of the order parameter is relatively diffuse in real space. By contrast, at $\nu=0.8$, vortices preferentially penetrate locally weak superconducting regions centered on vertex-A ($Z_i=8$) sites, and the field dependence of the spatially averaged order parameter exhibits a pronounced plateau-like regime accompanied by vortex clustering. The fact that these features are more prominent at the filling with a stronger enhancement of $B_{{\mathrm c}2}$ suggests that the field response is closely related to the intrinsic spatial inhomogeneity of the quasiperiodic superconducting state.

Indeed, our self-consistent BdG solutions exhibit a spatially modulated pairing amplitude $|\Delta_i|$, with robust superconducting regions coexisting with weaker ones, as shown in Figs.~\ref{fig:OP_T=0.01_n=0.5} and \ref{fig:OP_T=0.01_n=0.8}. This is also consistent with the discrepancy discussed in Sec.~\ref{subsec:Bc2} between the $B_{{\mathrm c}2}$ estimated from the effective coherence length extracted from isolated-vortex profiles and the numerically determined $B_{{\mathrm c}2}$. 
Taken together, these results suggest that quasiperiodicity-induced inhomogeneity can substantially modify both the magnitude and the temperature dependence of the orbital upper critical field. 
%\red{Since Zeeman pair breaking is not included in the present model, the calculated $B_{{\mathrm c}2}$ cannot be directly compared with the Pauli limit. Nevertheless, particularly at \(\nu=0.8\), the strong enhancement of $B_{{\mathrm c}2}$ and the absence of low-$T$ saturation are qualitatively reminiscent of the anomalous $B_{{\mathrm c}2}(T)$ behavior reported in the dodecagonal quasicrystal superconductor Ta\(_{1.6}\)Te \cite{Tokumoto2024,Terashima2024}.}
%Interestingly, a qualitatively similar absence of low-$T$ saturation and an increase of $B_{{\mathrm c}2}(T)$ upon cooling have been reported in the dodecagonal quasicrystal superconductor Ta$_{1.6}$Te \cite{Tokumoto2024,Terashima2024}.

The strong enhancement of $B_{{\mathrm c}2}$ and the absence of low-$T$ saturation, particularly at $\nu=0.8$, are qualitatively reminiscent of the anomalous $B_{{\mathrm c}2}(T)$ behavior reported in the dodecagonal quasicrystal superconductor Ta$_{1.6}$Te \cite{Tokumoto2024,Terashima2024}.
In addition, $B_{{\mathrm c}2}(T=0)\simeq0.76$ at $\nu=0.8$ is about twice the Pauli limit estimated as $1.84T_{\mathrm c}\simeq0.37$, in good agreement with the experimental observation \cite{Terashima2024}.
Although there are obvious differences in the crystal and electronic structures between the present model and Ta$_{1.6}$Te, and the Zeeman effect is not included in the calculation, our results suggest that quasiperiodicity may contribute to the anomalous behavior observed in Ta$_{1.6}$Te. This orbital mechanism could complement the effects of spin-orbit interactions on the spin response discussed in Ref.~\cite{Tada2025}.

%%%%%%%%%%%%%%%%%%%%%%%%%%%%%%%%%%%%%%%%%%%%%%%%%%%%%%%%%
\section{Conclusion}
\label{sec:summary}

\begin{figure*}[t]
    \centering
    \includegraphics[width=\linewidth]{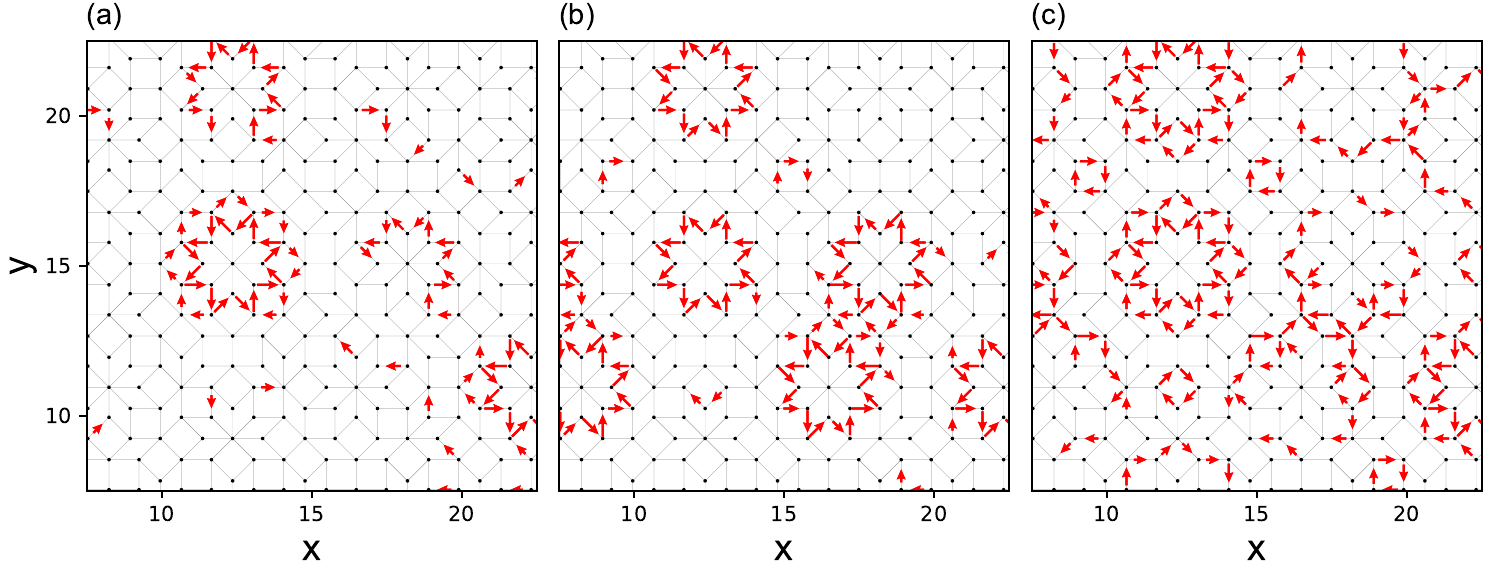}
    \caption{Real-space supercurrent distribution at $T=0.01$ for $\nu=0.8$. Gray lines and black dots denote the lattice links and sites, respectively. Red arrows indicate the direction and magnitude of the supercurrent; in each panel, the current magnitude is normalized by its maximum value, and only currents larger than one half of the panel-wise maximum are shown. Panels (a), (b), and (c) correspond to $B\simeq 0.5B_{\mathrm{c}2}$ ($N_\phi=65$), $0.7B_{\mathrm{c}2}$ ($N_\phi=90$), and $0.9B_{\mathrm{c}2}$ ($N_\phi=115$), respectively. For visibility, only the central $15\times 15$ region is shown in each panel.}
    \label{fig:supercurrent}
\end{figure*}

In this work, we investigated vortex states and the orbital upper critical field in an $s$-wave superconducting Ammann-Beenker approximant by self-consistently solving the BdG equations in a uniform magnetic field perpendicular to the quasiperiodic plane.
By analyzing the spatial distribution of the superconducting order parameter in both real and perpendicular spaces, and by comparing the results with those for a square lattice, we examined how quasiperiodicity modifies the collective organization of vortices and the destruction of superconductivity in the high-field regime.

We found that the high-field vortex response depends strongly on the filling. At $\nu=0.5$, the suppression of the order parameter remains relatively diffuse in real space, although the perpendicular-space representation reveals a clear dependence on the local vertex environment. 
By contrast, at $\nu=0.8$, vortices preferentially penetrate weak-pairing regions centered on vertex-A ($Z_i=8$) sites, and the high-field vortex organization is strongly influenced by the local geometry.
%By contrast, at $\nu=0.8$, vortices preferentially penetrate weak-pairing regions centered on vertex-A ($Z_i=8$) sites, and the resulting vortex distribution reflects the eightfold quasiperiodic organization of these local environments. 
%The persistence of the same qualitative features under open boundary conditions, including a finite cluster without global eightfold symmetry, further indicates that this spatial organization is associated primarily with the local geometry rather than with the boundary conditions or the global shape of the finite system.
%The persistence of these features under open boundary conditions, including for a finite cluster without global eightfold symmetry, further indicates that they originate primarily from the local quasiperiodic geometry rather than from the boundary conditions or the global shape of the finite system.
We also confirmed that the orbital upper critical field of the AB approximant is enhanced relative to that of the square lattice at both fillings. This enhancement is particularly pronounced for $\nu=0.8$ near zero temperature, where the low-field values of the spatially averaged order parameter are nearly identical for the two lattices. 
%The winding-number analysis shows that elementary vortices form densely packed clusters within locally weak structural regions, with intervortex separations much smaller than the effective vortex-core size estimated from isolated-vortex profiles. Consistently, the spatial fraction of strongly suppressed superconducting regions changes only weakly over the plateau regime, even as the magnetic flux continues to increase. Together, these observations support the interpretation that additional vortices can be accommodated within regions where superconductivity is already strongly weakened, allowing the surrounding, more robust superconducting regions to remain largely intact. 
At this filling, $\bar\Delta$ exhibits a pronounced plateau-like regime accompanied by vortex clusters in locally weak superconducting regions. The winding-number analysis and the spatial evolution of the suppressed regions support the view that additional vortices can be accommodated within regions where superconductivity is already weakened, providing a microscopic interpretation for the plateau-like behavior of $\bar{\Delta}$.
%This provides a microscopic interpretation of the plateau-like behavior of $\bar{\Delta}$ and its association with the enhanced $B_{c2}$.

Finally, the enhancement of the upper critical field persists over the temperature range up to the zero-field $T_{\mathrm c}$. At both fillings, $B_{{\mathrm c}2}(T)$ of the AB approximant remains larger than that of the square lattice, with the difference being particularly large at $\nu=0.8$. At this filling, $B_{{\mathrm c}2}(T)$ continues to increase upon cooling toward zero temperature without showing a clear tendency toward low-temperature saturation.
Taken together, our results show that the intrinsic spatial inhomogeneity generated by quasiperiodicity can qualitatively modify vortex organization and substantially affect both the magnitude and the temperature dependence of the orbital upper critical field. 
%The qualitatively similar low-temperature behavior 
A similar absence of low-temperature saturation in $B_{{\mathrm c}2}(T)$ reported in the dodecagonal quasicrystal superconductor Ta$_{1.6}$Te suggests that these effects may be relevant to superconducting quasicrystals more broadly.

%%%%%%%%%%%%%%%%%%%%%%%%%%%%%%%%%
\begin{acknowledgments}
The authors greatly appreciate inspiring discussions with Yuki Tokumoto and Yasuhiro Tada. 
N. T. is supported by JSPS KAKENHI Grant No. JP24K00585, JP25H01397, JP25H01400, J250603033.
S. S. is supported by JSPS KAKENHI Grant No. JP25H01397, JP25H01398, JP25K24854, and JP25K24855.
M. I. is supported by JSPS KAKENHI Grant No. JP26K07000.
\end{acknowledgments}

\section*{Data Availability}
The numerical data and code are available upon reasonable request from the corresponding authors.

\appendix

%%%%%%%%%%%%%%%%%%%%%%%%%%%%%%%%%%%%%%%%%%%%%%%%%%%%%%%%%%%%%%%%%%%%%%
\section{Supercurrent near vortex cores}
\label{app:supercurrent}

In the BdG formalism, the local supercurrent flowing from site $j$ to $i$ is given by
\begin{align}
\label{eq:super-J}
    \hat{\bm j}_{i,j} & = -\frac{\partial \hat{\cal H}({\bm A})}{\partial {\bm A}} \notag \\ & = 2t ~ \mathrm{Im}\!\left( e^{-{\rm i}\theta_{i,j}} \sum_{\sigma} \hat{c}^{\dagger}_{i\sigma}\hat{c}_{j\sigma} \right) {\bm r}_{i,j},
\end{align}
where ${\bm r}_{i,j}={\bm r}_i-{\bm r}_j$ is the relative vector.
The expectation value of $\hat{c}^{\dagger}_{i \sigma}\hat{c}_{j \sigma}$ in Eq.~\eqref{eq:super-J} is obtained from the eigenstates of the BdG equation as
\begin{equation}
\centering
\begin{gathered}
    \langle \hat{c}^{\dagger}_{i \uparrow}\hat{c}_{j \uparrow} \rangle
    = \sum_{\epsilon}u^{*}_{\epsilon}({\bm r}_{i})u_{\epsilon}({\bm r}_{j})f(E_{\epsilon}), \\
    \langle \hat{c}^{\dagger}_{i \downarrow}\hat{c}_{j \downarrow} \rangle
    = \sum_{\epsilon}v_{\epsilon}({\bm r}_{i})v^{*}_{\epsilon}({\bm r}_{j})f(-E_{\epsilon}).
\end{gathered}
\end{equation}

Figure~\ref{fig:supercurrent} shows the real-space distribution of the supercurrent at $T=0.01$ for $\nu=0.8$.
Panels (a), (b), and (c) correspond to $B\simeq 0.5B_{\mathrm{c}2}$ ($N_\phi=65$), $0.7B_{\mathrm{c}2}$ ($N_\phi=90$), and $0.9B_{\mathrm{c}2}$ ($N_\phi=115$), respectively.
In each panel, the supercurrent exhibits a circulating pattern around the local clusters centered on sites with $Z_i=8$.
This behavior is consistent with Fig.~\ref{fig:OP_T=0.01_n=0.8}, which shows that vortices preferentially penetrate the same clusters; consequently, the supercurrent circulates around them.
In panel (a), the circulating supercurrent flow appears to be limited to specific local clusters or some connected plaquettes. In panel (b), when local clusters centered on sites with $Z_i=8$ are located close to one another, one can clearly see the supercurrent flowing between the clusters. In panel (c), the circulating flow is no longer limited to isolated regions; instead,
%the supercurrent flows through the inter-cluster channels, 
a large supercurrent flows between clusters, forming a connected network of current paths.

%%%%%%%%%%%%%%%%%%%%%%%%%%%%%%%%%%%%%%%%%%%%%%%%%%%%%%%%%%%%%%%%%%%%%%
\section{Suppressed superconducting regions in the broadest plateau regime}
\label{app:plateau_suppressed_regions}

\begin{figure}[t]
    \centering
    \includegraphics[width=\linewidth]{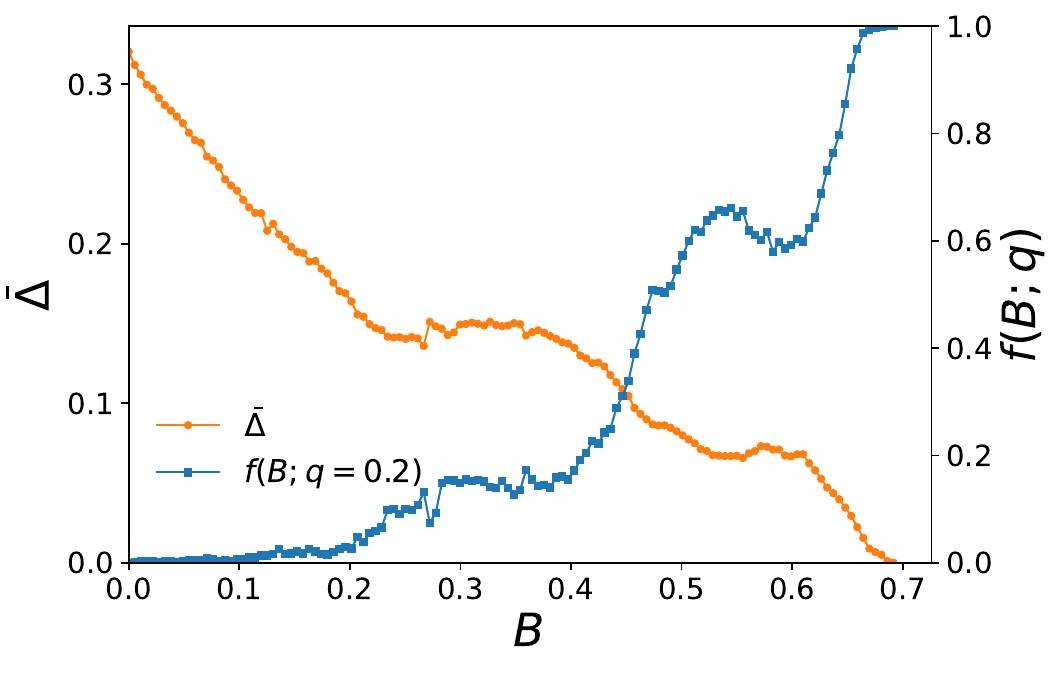}
    \caption{Comparison of the magnetic-field dependence of the spatially averaged order parameter $\bar{\Delta}$ (left axis) and the fraction $f(B;q=0.2)$ (right axis) of sites whose local amplitude is reduced below $20\%$ of its zero-field value, at $\nu=0.8$ and $T=0.01$.}
    \label{fig:op_suppression}
\end{figure}

To further characterize the plateau-like behavior discussed in Sec.~\ref{subsec:Bc2}, we examine the magnetic-field dependence of the spatial extent of the strongly suppressed superconducting regions.
We introduce the site-dependent normalized amplitude
\begin{equation}
    r_i(B) = \frac{|\Delta_i(B)|}{|\Delta_i(0)|},
\label{eq:local_suppression_ratio}
\end{equation}
where $|\Delta_i(0)|$ is the zero-field amplitude at the same site. Therefore $r_i(B)$ measures the field-induced reduction of the local pairing amplitude relative to its zero-field value.
Using Eq.~\eqref{eq:local_suppression_ratio}, we define the fraction of strongly suppressed sites as
\begin{equation}
    f(B;q) =
    \frac{1}{N}\sum^N_{i=1}\Theta\left(q-r_i(B)\right),
\label{eq:suppressed_fraction}
\end{equation}
where $\Theta(x)$ is the step function. 
The quantity $f(B;q)$ therefore represents the ratio of sites whose local amplitude is reduced below a fraction $q$ of its zero-field value.

Figure \ref{fig:op_suppression} shows $f(B;q)$ for $q=0.2$ at $\nu=0.8$ and $T=0.01$. 
This provides a criterion for identifying sites at which the local pairing amplitude is strongly suppressed.
At low fields, $f(B;0.2)$ changes only weakly. %$f(B;0.2)$ increases as vortices enter the system and suppress the pairing amplitude in their vicinity. 
In the first plateau regime of $\bar{\Delta}(B)$, i.e., $0.2\lesssim B\lesssim0.4$, $f(B;0.2)$ also changes only weakly even as the number of flux quanta increases.
%the increase of $f(B;0.2)$ becomes substantially weaker. 
After the system leaves this regime, $f(B;0.2)$ increases rapidly, indicating that the strongly suppressed-$|\Delta_i|$ regions begin to extend over a larger fraction of the system. Thus, although the number of flux quanta continues to increase within the plateau regime, the additional flux does not produce a proportional increase in the spatial extent of the strongly suppressed regions. 
%Instead, the suppression remains concentrated predominantly in regions \red{where the superconductivity has} already been substantially weakened.
This behavior is consistent with the suppression remaining concentrated predominantly in regions where the superconductivity has already been substantially weakened.

%This behavior is also consistent with the vortex configurations obtained from the winding-number analysis. 
The vortex configurations obtained from the winding-number analysis further support this interpretation.
In the relevant field range, neighboring vortices tend to occupy the same weak-$|\Delta_i|$ regions, resulting in substantial spatial overlap of the regions surrounding the vortex cores. 
The present analysis characterizes only the spatial fraction of strongly suppressed sites and does not by itself establish a direct causal relation between vortex clustering and the plateau.
%alone does not determine the connectivity or detailed geometry of the vortex configurations and therefore does not establish a direct causal relation between vortex clustering and the plateau. 
Nevertheless, when combined with the winding-number analysis in Sec.~\ref{subsec:Bc2}, 
it supports the interpretation that, within the plateau regime, the quasiperiodic system can accommodate additional flux without immediately expanding the total region in which superconductivity is strongly suppressed.

\begin{figure*}[t]
    \centering
    \includegraphics[width=\textwidth]{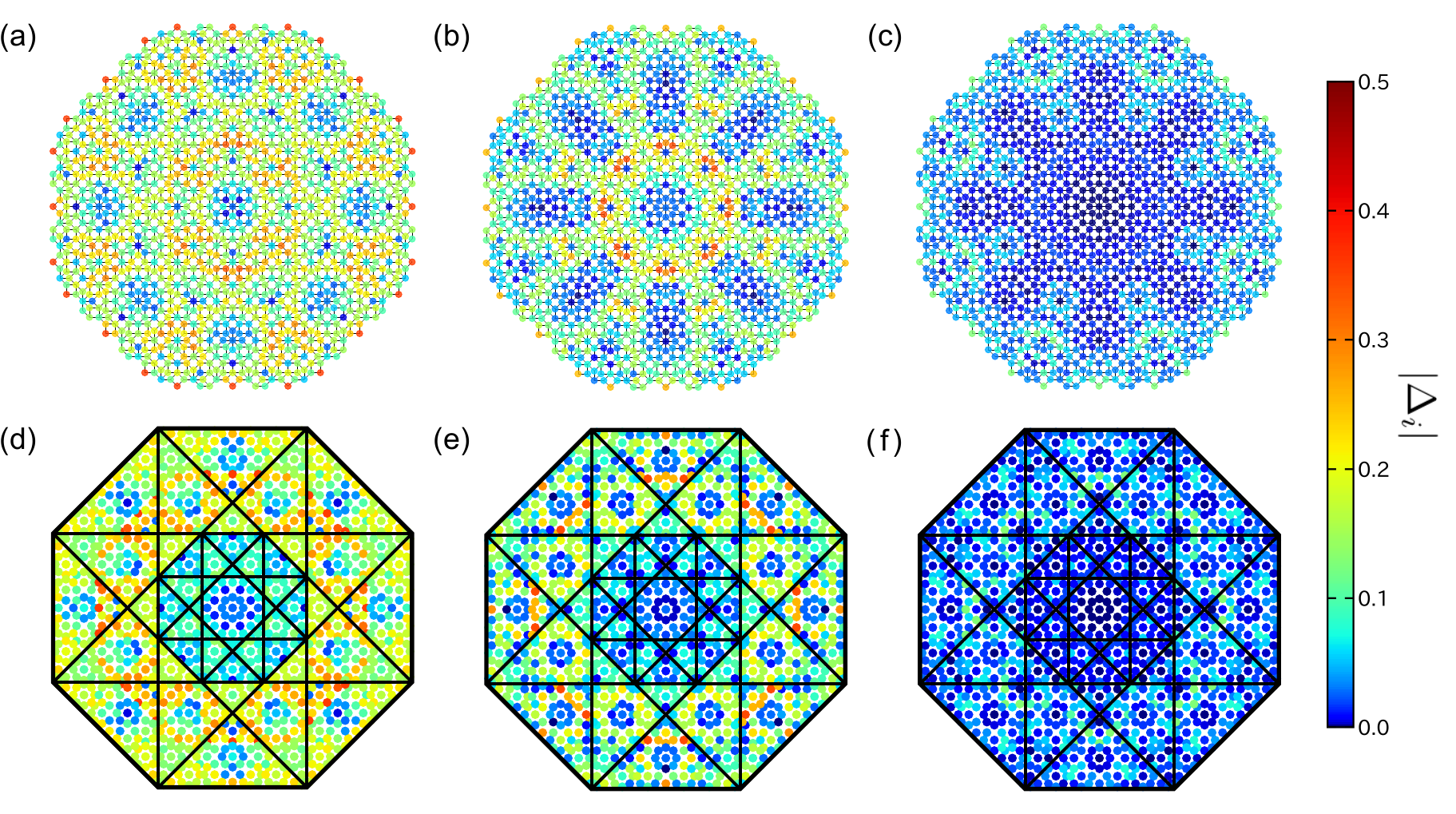}
    \caption{Spatial distributions of the order parameter amplitude $|\Delta_i|$ in real space [panels (a)-(c)] and perpendicular space [panels (d)-(f)] for the eightfold-symmetric Ammann-Beenker cluster under open boundary conditions at $T=0.01$ and $\nu=0.8$. The left, middle, and right panels correspond to fields chosen to approximately match the total flux of the periodic-boundary calculations in Fig.~\ref{fig:OP_T=0.01_n=0.8} at $B\approx0.5B_{\mathrm{c}2}$, $0.7B_{\mathrm{c}2}$, and $0.9B_{\mathrm{c}2}$, respectively.}
    \label{fig:OBC_symm}
\end{figure*}

%%%%%%%%%%%%%%%%%%%%%%%%%%%%%%%%%%%%%%%%%%%%%%%%%%%%%%%%%%%%%%%%%%%%%%
\section{Open boundary conditions}
\label{app:OBCs}

\begin{figure*}[t]
    \centering
    \includegraphics[width=\textwidth]{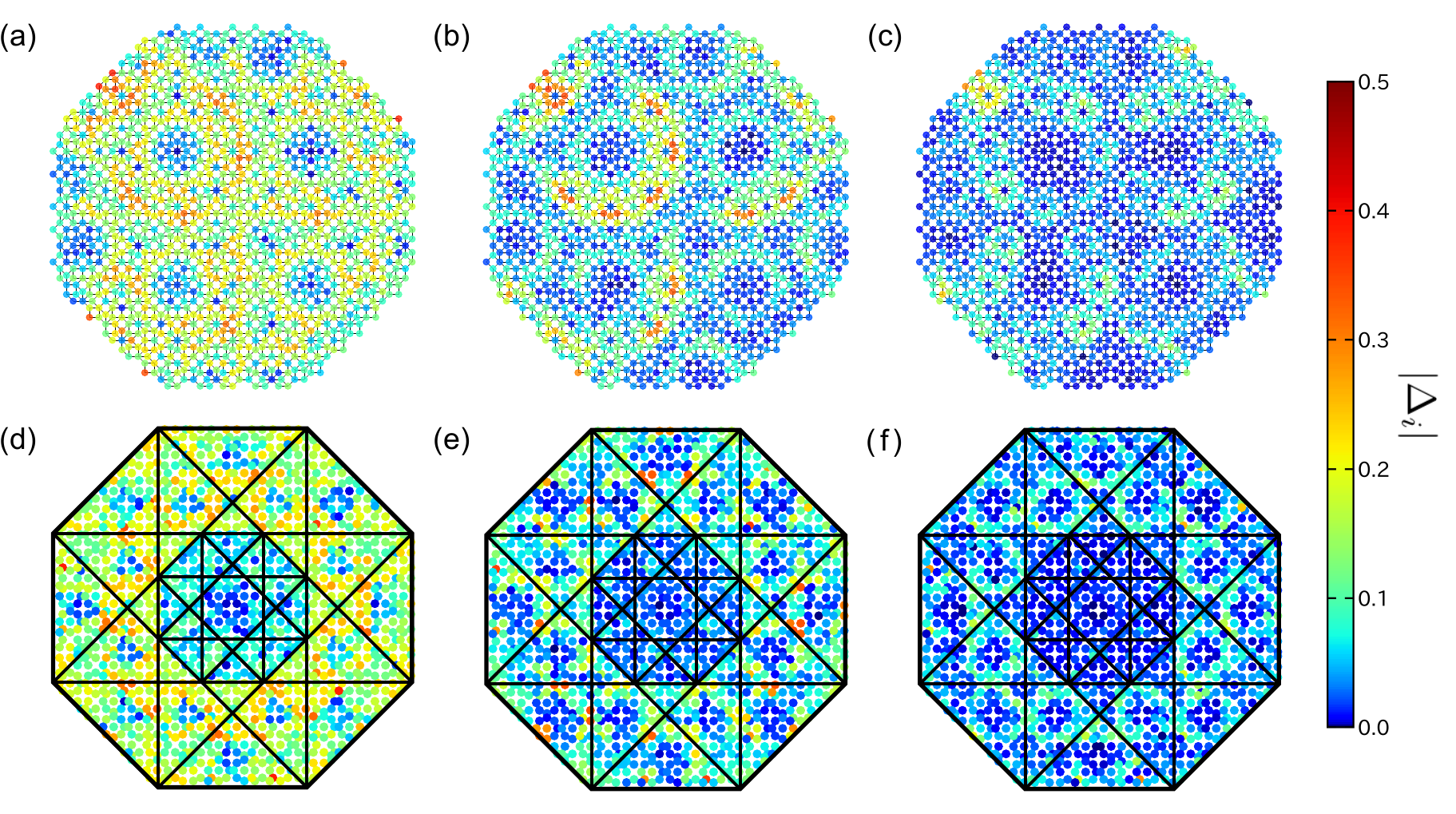}
    \caption{Spatial distributions of the order parameter amplitude $|\Delta_i|$ in real space [panels (a)-(c)] and perpendicular space [panels (d)-(f)] for the nonsymmetric Ammann-Beenker cluster under open boundary conditions at $T=0.01$ and $\nu=0.8$. The left, middle, and right panels correspond to fields chosen to approximately match the total flux of the periodic-boundary calculations in Fig.~\ref{fig:OP_T=0.01_n=0.8} at $B\approx0.5B_{\mathrm{c}2}$, $0.7B_{\mathrm{c}2}$, and $0.9B_{\mathrm{c}2}$, respectively.}
    \label{fig:OBC_nonsymm}
\end{figure*}

In the main text, we employed a square approximant with periodic boundary conditions. To verify that the vortex structures discussed above are not artifacts of either the boundary condition or the global symmetry of the approximant, we performed additional self-consistent BdG calculations under open boundary conditions.
We consider two finite clusters of the AB tiling generated using the cut-and-project method \cite{Beenker1982, Duneau1985}. 
The first cluster has exact eightfold rotational symmetry about its center and contains $N=1433$ sites with an area $A=1128.029$, whereas the second is a generic cluster without global eightfold symmetry and contains $N=1439$ sites with an area $A=1132.271$. 
Here, the tile-edge length is set to unity, so that the area of a square is one.
The latter is particularly useful for distinguishing effects arising from the local quasiperiodic geometry from those imposed by the overall shape of the finite system. 
We use the Landau gauge as in the periodic-boundary calculation. Under open boundary conditions, however, the total flux through the finite cluster is not constrained by the condition in Eq.~\eqref{eq:magnetic_field}, and $B$ can therefore be chosen arbitrarily.
For comparison with the periodic approximant, we choose the magnetic fields such that the total flux $\Phi=BA$ through each finite cluster is approximately equal to that used for the corresponding periodic-boundary calculation in Fig.~\ref{fig:OP_T=0.01_n=0.8}.

Figures \ref{fig:OBC_symm} and \ref{fig:OBC_nonsymm} show the spatial distributions of the order-parameter amplitude $\lvert\Delta_i\rvert$ at $T=0.01$ and $\nu=0.8$ for the eightfold-symmetric and nonsymmetric AB clusters, respectively. 
In both clusters, the real-space distributions evolve with increasing magnetic field in the same qualitative manner as those obtained for the square approximant with periodic boundary conditions: superconductivity is initially suppressed in spatially separated regions, the suppressed regions broaden and become increasingly connected at intermediate fields, and only limited superconducting regions remain at the highest fields shown. %survive near the upper critical field. 
The corresponding perpendicular-space distributions also exhibit a similar field evolution, with superconductivity being preferentially depleted in the same local-environment sectors as in the periodic approximant. 
Although the detailed patterns are modified by finite-size and boundary effects, particularly for the nonsymmetric cluster, the characteristic qualitative features of the high-field organization persist.
This agreement indicates that the observed spatial organization is not imposed by the boundary conditions or by the global eightfold symmetry of the finite sample, and instead supports its association with the local quasiperiodic geometry.

\begin{figure}[t]
    \centering
    \includegraphics[width=\linewidth]{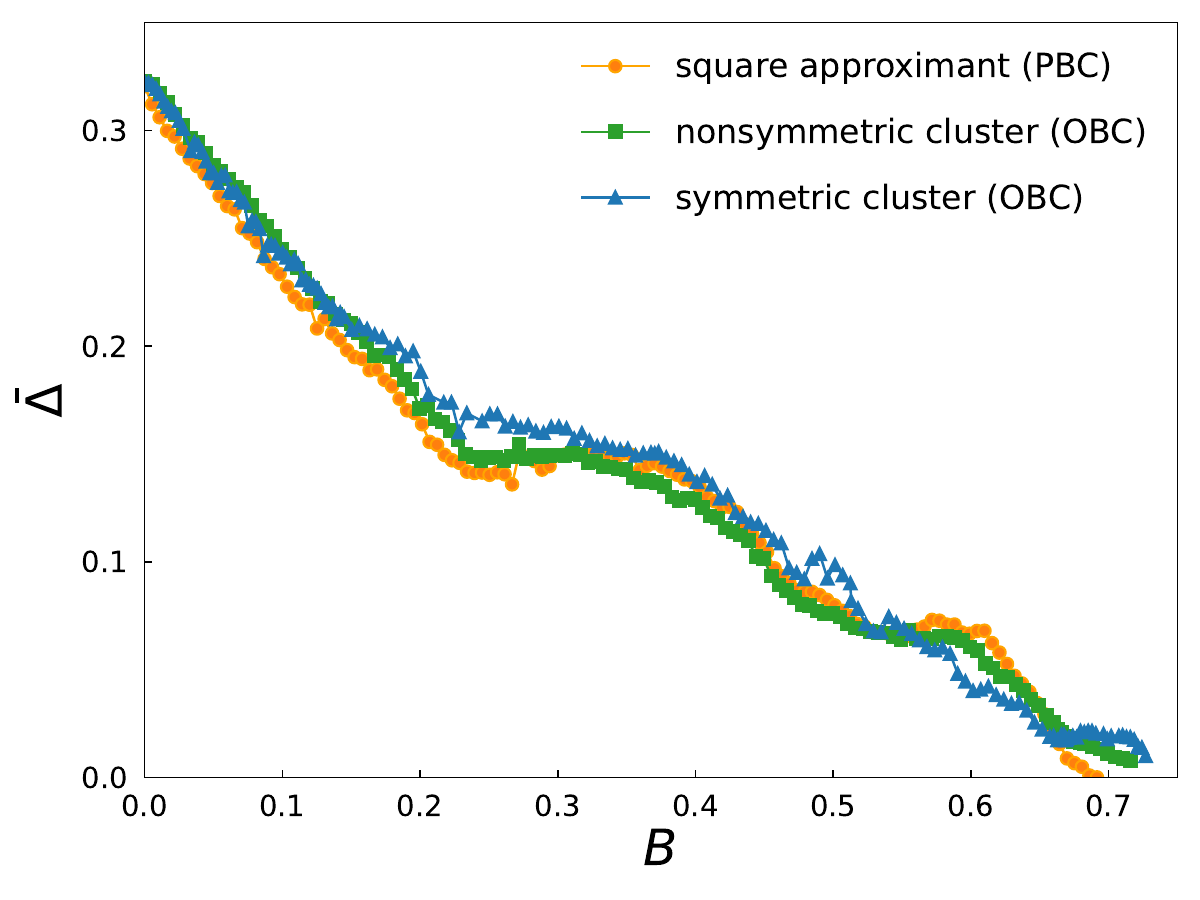}
    \caption{Magnetic-field dependence of the spatially averaged order-parameter amplitude $\bar\Delta$ at $\nu=0.8$ and $T=0.01$.
    The orange circles show the results for the square approximant with periodic boundary conditions, while the green squares and blue triangles show the results for the nonsymmetric and eightfold-symmetric Ammann-Beenker clusters with open boundary conditions, respectively.}
    \label{fig:comparison_averageOP}
\end{figure}

Figure \ref{fig:comparison_averageOP} further compares the magnetic-field dependence of the spatially averaged amplitude $\bar\Delta$ at $\nu=0.8$ and $T=0.01$ for the periodic approximant and for the nonsymmetric and eightfold-symmetric Ammann-Beenker clusters with open boundary conditions. Despite the quantitative differences associated with finite-size and boundary effects, the three curves exhibit very similar field dependence. In particular, the initial decrease of $\bar\Delta$, the broad plateau-like regime at intermediate fields, the subsequent suppression at higher fields, and a comparable overall scale of the upper critical field are observed in both open clusters. 
The agreement with the periodic approximant again indicates that these features are not artifacts of the boundary conditions, while the similarity between the symmetric and nonsymmetric clusters shows that they are not imposed by the global eightfold symmetry of the system boundary.

These results show that the main qualitative conclusions are robust against the choice of boundary conditions. In particular, the preferential suppression of superconductivity in locally weak, vertex-A-centered regions and the associated field evolution persist under open boundary conditions. The plateau-like field dependence of the spatially averaged order parameter is also reproduced in both open clusters. 
Moreover, the results for the nonsymmetric cluster show that these features do not require an eightfold-symmetric sample boundary, supporting the interpretation that they are associated primarily with the local quasiperiodic geometry.

%%%%%%%%%%%%%%%%%%%%%%%%%%%%%%%%%
\bibliography{ref_new}

@article{Shechtman1984,
  title = {Metallic {P}hase with {L}ong-{R}ange {O}rientational {O}rder and {N}o {T}ranslational {S}ymmetry},
  author = {Shechtman, D. and Blech, I. and Gratias, D. and Cahn, J. W.},
  journal = {Phys. Rev. Lett.},
  volume = {53},
  issue = {20},
  pages = {1951--1953},
  numpages = {0},
  year = {1984},
  month = {Nov},
  publisher = {American Physical Society},
  doi = {10.1103/PhysRevLett.53.1951},
  url = {https://link.aps.org/doi/10.1103/PhysRevLett.53.1951}
}

@article{Levine1984,
  title = {Quasicrystals: {A} {N}ew {C}lass of {O}rdered {S}tructures},
  author = {Levine, Dov and Steinhardt, Paul Joseph},
  journal = {Phys. Rev. Lett.},
  volume = {53},
  issue = {26},
  pages = {2477--2480},
  numpages = {0},
  year = {1984},
  month = {Dec},
  publisher = {American Physical Society},
  doi = {10.1103/PhysRevLett.53.2477},
  url = {https://link.aps.org/doi/10.1103/PhysRevLett.53.2477}
}

@article{Kamiya2018,
  author = {Kamiya, K and Takeuchi, T and Kabeya, N and Wada, N and Ishimasa, T and Ochiai, A and Deguchi, K and Imura, K and Sato, N K},
  doi = {10.1038/s41467-017-02667-x},
  issn = {2041-1723},
  journal = {Nat. Comm.},
  number = {1},
  pages = {154},
  title = {{Discovery of superconductivity in quasicrystal}},
  url = {https://doi.org/10.1038/s41467-017-02667-x},
  volume = {9},
  year = {2018}
}

@article{Tokumoto2024,
  title = {Superconductivity in a van der {W}aals layered quasicrystal},
  author = {Tokumoto, Yuki and Hamano, Kotaro and Nakagawa, Sunao and Kamimura, Yasushi and Suzuki, Shintaro and Tamura, Ryuji and Edagawa, Keiichi},
  journal = {Nat. Comm.},
  volume = {15},
  number = {1},
  pages = {1529},
  year = {2024},
  publisher = {Nature Publishing Group UK London},
  url = {https://www.nature.com/articles/s41467-024-45952-2}
}

@article{Abrikosov1957,
  title = {The magnetic properties of superconducting alloys},
  author = {Abrikosov, Alexei Alexeyevich},
  journal = {J. Phys. Chem. Sol.},
  volume = {2},
  number = {3},
  pages = {199--208},
  year = {1957},
  publisher = {Elsevier},
  url = {https://www.sciencedirect.com/science/article/abs/pii/0022369757900835}
}

@article{Sakai2017,
  title = {Superconductivity on a quasiperiodic lattice: {E}xtended-to-localized crossover of {C}ooper pairs},
  author = {Sakai, Shiro and Takemori, Nayuta and Koga, Akihisa and Arita, Ryotaro},
  journal = {Phys. Rev. B},
  volume = {95},
  issue = {2},
  pages = {024509},
  numpages = {5},
  year = {2017},
  month = {Jan},
  publisher = {American Physical Society},
  doi = {10.1103/PhysRevB.95.024509},
  url = {https://journals.aps.org/prb/abstract/10.1103/PhysRevB.95.024509}
}

@article{Takemori2020,
  title = {Physical properties of weak-coupling quasiperiodic superconductors},
  author = {Takemori, Nayuta and Arita, Ryotaro and Sakai, Shiro},
  journal = {Phys. Rev. B},
  volume = {102},
  issue = {11},
  pages = {115108},
  numpages = {8},
  year = {2020},
  month = {Sep},
  publisher = {American Physical Society},
  doi = {10.1103/PhysRevB.102.115108}
}

@article{Fukushima2023,
  title = {Supercurrent distribution in real-space and anomalous paramagnetic response in a superconducting quasicrystal},
  author = {Fukushima, Takumi and Takemori, Nayuta and Sakai, Shiro and Ichioka, Masanori and Jagannathan, Anuradha},
  journal = {Phys. Rev. Res.},
  volume = {5},
  issue = {4},
  pages = {043164},
  numpages = {13},
  year = {2023},
  month = {Nov},
  publisher = {American Physical Society},
  doi = {10.1103/PhysRevResearch.5.043164},
  url = {https://link.aps.org/doi/10.1103/PhysRevResearch.5.043164}
}

@book{Beenker1982,
  title = {Algebraic theory of non-periodic tilings of the plane by two simple building blocks : a square and a rhombus},
  author = {F.P.M. Beenker},
  year = {1982},
  series = {EUT report. WSK, Dept. of Mathematics and Computing Science},
  publisher = {Eindhoven University of Technology},
  url = {https://research.tue.nl/en/publications/algebraic-theory-of-non-periodic-tilings-of-the-plane-by-two-simp/}
}

@article{Ammann1992,
  title = {Aperiodic tiles},
  author = {Ammann, Robert and Gr{\"u}nbaum, Branko and Shephard, Geoffrey C},
  journal = {Discrete \& Computational Geometry},
  volume = {8},
  number = {1},
  pages = {1--25},
  year = {1992},
  publisher = {Springer},
  url = {https://link.springer.com/article/10.1007/BF02293033}
}

@article{Jagannathan2004,
  title = {Quantum {S}pins and {Q}uasiperiodicity: {A} {R}eal {S}pace {R}enormalization {G}roup {A}pproach},
  author = {Jagannathan, A},
  journal = {Phys. {R}ev. {L}ett.},
  volume = {92},
  number = {4},
  pages = {047202},
  year = {2004},
  publisher = {APS},
  url = {https://journals.aps.org/prl/abstract/10.1103/PhysRevLett.92.047202}
}

@article{Jagannathan2005,
  title = {Ground state of a two-dimensional quasiperiodic quantum antiferromagnet},
  author = {Jagannathan, A.},
  journal = {Phys. Rev. B},
  volume = {71},
  issue = {11},
  pages = {115101},
  numpages = {14},
  year = {2005},
  month = {Mar},
  publisher = {American Physical Society},
  doi = {10.1103/PhysRevB.71.115101}
}

@article{Sakai2021,
  title = {Effect of {E}lectron-{E}lectron {I}nteractions on {M}etallic {S}tate in {Q}uasicrystals},
  author = {Sakai, Shiro and Koga, Akihisa},
  journal = {Mater. Trans.},
  volume = {62},
  number = {3},
  pages = {380--385},
  year = {2021},
  publisher = {The Japan Institute of Metals and Materials},
  url = {https://www.jstage.jst.go.jp/article/matertrans/62/3/62_MT-MB2020001/_article/-char/ja/}
}

@book{Peierls1997,
  author = {R. Peierls},
  title = {On the {T}heory of the {D}iamagnetism of {C}onduction {E}lectrons},
  booktitle = {Selected Scientific Papers of Sir Rudolf Peierls},
  chapter = {},
  pages = {},
  year = {1997},
  publisher = {World Scientific},
  doi = {10.1142/9789812795779_0010},
  url = {https://www.worldscientific.com/doi/abs/10.1142/9789812795779_0010}
}

@book{deGennes1999,
  author = {De Gennes, Pierre Gilles},
  title = {Superconductivity of Metals and Alloys},
  publisher = {Perseus},
  series = {Advanced book classics},
  year = {1999},
  url = {https://www.taylorfrancis.com/books/mono/10.1201/9780429497032/superconductivity-metals-alloys-de-gennes},
}

@article{Duneau1989,
  title={Approximants of quasiperiodic structures generated by the inflation mapping},
  author={Duneau, Michel},
  journal={J. Phys. A, Math. Gen.},
  volume={22},
  number={21},
  pages={4549},
  year={1989},
  publisher={IOP Publishing},
  url={https://iopscience.iop.org/article/10.1088/0305-4470/22/21/017/meta?casa_token=HoWsVt2Dl7UAAAAA:h0lAChZk0fFKMddstHUoN0FpKIMnvyn1xPqrifvuqEboAPkS54gJslRDAGMFbwh1xi_88-LOV3365cnWNBIpXNdDy18}
}

@article{Jagannathan2024,
  title={Properties of the {A}mmann-{B}eenker tiling and its square periodic approximants},
  author={Jagannathan, Anuradha and Duneau, Michel},
  journal={Isr. J. Chem.},
  volume={64},
  number={10-11},
  pages={e202300119},
  year={2024},
  publisher={Wiley Online Library},
  url={https://onlinelibrary.wiley.com/doi/full/10.1002/ijch.202300119}
}

@article{Araujo2019,
  title = {Conventional superconductivity in quasicrystals},
  author = {Ara\'ujo, Ronaldo N. and Andrade, Eric C.},
  journal = {Phys. Rev. B},
  volume = {100},
  issue = {1},
  pages = {014510},
  numpages = {10},
  year = {2019},
  month = {Jul},
  publisher = {American Physical Society},
  doi = {10.1103/PhysRevB.100.014510},
  url = {https://journals.aps.org/prb/abstract/10.1103/PhysRevB.100.014510}
}

@article{Nagai2022,
  title = {Intrinsic vortex pinning in superconducting quasicrystals},
  author = {Nagai, Yuki},
  journal = {Phys. Rev. B},
  volume = {106},
  issue = {6},
  pages = {064506},
  numpages = {10},
  year = {2022},
  month = {Aug},
  publisher = {American Physical Society},
  doi = {10.1103/PhysRevB.106.064506},
  url = {https://link.aps.org/doi/10.1103/PhysRevB.106.064506}
}

@article{Fukushima2023proceeding,
  title={Supercurrent {D}istribution on {A}mmann-{B}eenker {S}tructure},
  author={Fukushima, Takumi and Takemori, Nayuta and Sakai, Shiro and Ichioka, Masanori and Jagannathan, Anuradha},
  journal={J. Phys.: Conf. Ser.},
  volume={2461},
  number={1},
  pages={012014},
  year={2023},
  organization={IOP Publishing},
  url={https://iopscience.iop.org/article/10.1088/1742-6596/2461/1/012014/meta}
}

@article{Bardeen1957,
  title={Theory of {S}uperconductivity},
  author={Bardeen, John and Cooper, Leon N and Schrieffer, John Robert},
  journal={Phys. Rev.},
  volume={108},
  number={5},
  pages={1175},
  year={1957},
  publisher={APS},
  url={https://journals.aps.org/pr/abstract/10.1103/PhysRev.108.1175}
}

@article{Liu2023,
  title={Unconventional superfluidity of superconductivity on {P}enrose lattice},
  author={Liu, Yu-Bo and Shao, Zhi-Yan and Cao, Ye and Yang, Fan},
  journal={Sci. China-Phys. Mech. Astron.},
  volume={66},
  number={9},
  pages={290312},
  year={2023},
  publisher={Springer},
  url={https://link.springer.com/article/10.1007/s11433-023-2139-0}
}

@article{Liu2022,
  title={Cooper instability and superconductivity of the {P}enrose lattice},
  author={Liu, Yu-Bo and Hao, Juan-Juan and Zhang, Yongyou and Cao, Ye and Chen, Wei-Qiang and Yang, Fan},
  journal={Sci. China-Phys. Mech. Astron.},
  volume={65},
  number={8},
  pages={287411},
  year={2022},
  publisher={Springer},
  url={https://link.springer.com/article/10.1007/s11433-021-1877-5}
}

@book{Kita2015,
  title={Statistical Mechanics of Superconductivity},
  author={Kita, Takafumi},
  year={2015},
  publisher={Springer},
  url={https://link.springer.com/book/10.1007/978-4-431-55405-9}
}

@book{Tinkham2004,
  title={Introduction to {S}uperconductivity},
  author={Tinkham, Michael},
  year={2004},
  publisher={Courier Corporation},
  url={https://books.google.co.jp/books?hl=ja&lr=lang_ja|lang_en&id=VpUk3NfwDIkC&oi=fnd&pg=PP1&dq=introduction+to+superconductivity+tinkham&ots=xueKt5uQSl&sig=cSUc8KRd9sxjDHEIFu9DlpMruPg#v=onepage&q&f=false}
}

@article{Terashima2024,
  title={Anomalous upper critical field in the quasicrystal superconductor $\rm{{T}a_{1.6}{T}e}$},
  author={Terashima, Taichi and Tokumoto, Yuki and Hamano, Kotaro and Konoike, Takako and Kikugawa, Naoki and Edagawa, Keiichi},
  journal={npj Quantum Mater.},
  volume={9},
  number={1},
  pages={56},
  year={2024},
  publisher={Nature Publishing Group UK London},
  url={https://www.nature.com/articles/s41535-024-00669-9}
}

@article{Kohmoto1983,
  title = {Localization Problem in One Dimension: Mapping and Escape},
  author = {Kohmoto, Mahito and Kadanoff, Leo P. and Tang, Chao},
  journal = {Phys. Rev. Lett.},
  volume = {50},
  issue = {23},
  pages = {1870--1872},
  numpages = {0},
  year = {1983},
  month = {Jun},
  publisher = {American Physical Society},
  doi = {10.1103/PhysRevLett.50.1870},
  url = {https://link.aps.org/doi/10.1103/PhysRevLett.50.1870}
}

@article{Ostlund1983,
  title = {One-{D}imensional {S}chr\"odinger {E}quation with an {A}lmost {P}eriodic {P}otential},
  author = {Ostlund, Stellan and Pandit, Rahul and Rand, David and Schellnhuber, Hans Joachim and Siggia, Eric D.},
  journal = {Phys. Rev. Lett.},
  volume = {50},
  issue = {23},
  pages = {1873--1876},
  numpages = {0},
  year = {1983},
  month = {Jun},
  publisher = {American Physical Society},
  doi = {10.1103/PhysRevLett.50.1873},
  url = {https://link.aps.org/doi/10.1103/PhysRevLett.50.1873}
}

@article{Niu1986,
  title = {Renormalization-{G}roup {S}tudy of {O}ne-{D}imensional {Q}uasiperiodic {S}ystems},
  author = {Niu, Qian and Nori, Franco},
  journal = {Phys. Rev. Lett.},
  volume = {57},
  issue = {16},
  pages = {2057--2060},
  numpages = {0},
  year = {1986},
  month = {Oct},
  publisher = {American Physical Society},
  doi = {10.1103/PhysRevLett.57.2057},
  url = {https://link.aps.org/doi/10.1103/PhysRevLett.57.2057}
}

@article{Tsunetsugu1986,
  title={Eigenstates in 2-{D}imensional {P}enrose {T}iling},
  author={Tsunetsugu, Hirokazu and Fujiwara, Takeo and Ueda, Kazuo and Tokihiro, Tetsuji},
  journal={J. Phys. Soc. Jpn.},
  volume={55},
  number={5},
  pages={1420--1423},
  year={1986},
  publisher={The Physical Society of Japan},
  url={https://journals.jps.jp/doi/abs/10.1143/JPSJ.55.1420?journalCode=jpsj}
}

@article{Kohmoto1987,
  title = {Critical wave functions and a {C}antor-set spectrum of a one-dimensional quasicrystal model},
  author = {Kohmoto, Mahito and Sutherland, Bill and Tang, Chao},
  journal = {Phys. Rev. B},
  volume = {35},
  issue = {3},
  pages = {1020--1033},
  numpages = {0},
  year = {1987},
  month = {Jan},
  publisher = {American Physical Society},
  doi = {10.1103/PhysRevB.35.1020},
  url = {https://link.aps.org/doi/10.1103/PhysRevB.35.1020}
}

@article{Tokihiro1988,
  title = {Exact eigenstates on a two-dimensional {P}enrose lattice and their fractal dimensions},
  author = {Tokihiro, Tetsuji and Fujiwara, Takeo and Arai, Masao},
  journal = {Phys. Rev. B},
  volume = {38},
  issue = {9},
  pages = {5981--5987},
  numpages = {0},
  year = {1988},
  month = {Sep},
  publisher = {American Physical Society},
  doi = {10.1103/PhysRevB.38.5981},
  url = {https://link.aps.org/doi/10.1103/PhysRevB.38.5981}
}

@article{Tsunetsugu1991,
  title = {Electronic properties of the {P}enrose lattice. {I}. {E}nergy spectrum and wave functions},
  author = {Tsunetsugu, Hirokazu and Fujiwara, Takeo and Ueda, Kazuo and Tokihiro, Tetsuji},
  journal = {Phys. Rev. B},
  volume = {43},
  issue = {11},
  pages = {8879--8891},
  numpages = {0},
  year = {1991},
  month = {Apr},
  publisher = {American Physical Society},
  doi = {10.1103/PhysRevB.43.8879},
  url = {https://link.aps.org/doi/10.1103/PhysRevB.43.8879}
}

@article{Bloch1928,
  title={{\"U}ber die {Q}uantenmechanik der {E}lektronen in {K}ristallgittern},
  author={Bloch, Felix},
  journal={Z. Phys.},
  volume={52},
  number={7-8},
  pages={555--600},
  year={1928},
  publisher={Springer},
  url={https://link.springer.com/article/10.1007/BF01339455#article-info}
}

@book{Ashcroft1976,
  title={Solid {S}tate {P}hysics},
  author={Ashcroft, N.W. and Mermin, N.D.},
  isbn={9780030839931},
  lccn={lc74009772},
  series={HRW international editions},
  url={https://books.google.co.jp/books?id=oXIfAQAAMAAJ},
  year={1976},
  publisher={Holt, Rinehart and Winston}
}

@article{Anderson1958,
  title={Absence of {D}iffusion in {C}ertain {R}andom {L}attices},
  author={Anderson, Philip W},
  journal={Phys. Rev.},
  volume={109},
  number={5},
  pages={1492},
  year={1958},
  publisher={APS},
  url={https://journals.aps.org/pr/abstract/10.1103/PhysRev.109.1492}
}

@article{Lee1985,
  title={Disordered electronic systems},
  author={Lee, Patrick A and Ramakrishnan, Tiruppattur V},
  journal={Rev. Mod. Phys.},
  volume={57},
  number={2},
  pages={287},
  year={1985},
  publisher={APS},
  url={https://journals.aps.org/rmp/abstract/10.1103/RevModPhys.57.287}
}

@article{Matsudaira2026,
  title = {Unusual {NMR} signatures of superconductivity in the two-dimensional quasicrystal ${{(\mathrm{Ta}}_{0.95}{\mathrm{Cu}}_{0.05})}_{1.6}\mathrm{Te}$},
  author = {Matsudaira, H. and Kitagawa, S. and Ishida, K. and Tokumoto, Y. and Tomiyama, K. and Edagawa, K.},
  journal = {Phys. Rev. Mater.},
  volume = {10},
  issue = {6},
  pages = {064804},
  numpages = {7},
  year = {2026},
  month = {Jun},
  publisher = {American Physical Society},
  doi = {10.1103/kyh6-vt3q},
  url = {https://link.aps.org/doi/10.1103/kyh6-vt3q}
}

@book{Zhu2016,
  title={Bogoliubov-de {G}ennes {M}ethod and {I}ts {A}pplications},
  author={Zhu, Jian-Xin},
  year={2016},
  publisher={Springer},
  url={https://link.springer.com/book/10.1007/978-3-319-31314-6}
}

@article{Sakai2019,
  title = {Exotic pairing state in quasicrystalline superconductors under a magnetic field},
  author = {Sakai, Shiro and Arita, Ryotaro},
  journal = {Phys. Rev. Res.},
  volume = {1},
  issue = {2},
  pages = {022002},
  numpages = {5},
  year = {2019},
  month = {Sep},
  publisher = {American Physical Society},
  doi = {10.1103/PhysRevResearch.1.022002},
  url = {https://link.aps.org/doi/10.1103/PhysRevResearch.1.022002}
}

@article{Kleiner1964,
  title = {Bulk {S}olution of {G}inzburg-{L}andau {E}quations for {T}ype {II} {S}uperconductors: {U}pper {C}ritical {F}ield {R}egion},
  author = {Kleiner, W. H. and Roth, L. M. and Autler, S. H.},
  journal = {Phys. Rev.},
  volume = {133},
  issue = {5A},
  pages = {A1226--A1227},
  numpages = {0},
  year = {1964},
  month = {Mar},
  publisher = {American Physical Society},
  doi = {10.1103/PhysRev.133.A1226},
  url = {https://link.aps.org/doi/10.1103/PhysRev.133.A1226}
}

@article{Essmann1967,
title = {The direct observation of individual flux lines in type {II} superconductors},
journal = {Phys. Lett. A},
volume = {24},
number = {10},
pages = {526-527},
year = {1967},
issn = {0375-9601},
doi = {https://doi.org/10.1016/0375-9601(67)90819-5},
url = {https://www.sciencedirect.com/science/article/pii/0375960167908195},
author = {U. Essmann and H. Träuble}
}

@article{Hess1989,
  title = {Scanning-{T}unneling-{M}icroscope {O}bservation of the {A}brikosov {F}lux {L}attice and the {D}ensity of {S}tates near and inside a {F}luxoid},
  author = {Hess, H. F. and Robinson, R. B. and Dynes, R. C. and Valles, J. M. and Waszczak, J. V.},
  journal = {Phys. Rev. Lett.},
  volume = {62},
  issue = {2},
  pages = {214--216},
  numpages = {0},
  year = {1989},
  month = {Jan},
  publisher = {American Physical Society},
  doi = {10.1103/PhysRevLett.62.214},
  url = {https://link.aps.org/doi/10.1103/PhysRevLett.62.214}
}

@article{Wang1995,
  title = {Mixed-state quasiparticle spectrum for $d$-wave superconductors},
  author = {Wang, Yong and MacDonald, A. H.},
  journal = {Phys. Rev. B},
  volume = {52},
  issue = {6},
  pages = {R3876(R)--R3879(R)},
  numpages = {0},
  year = {1995},
  month = {Aug},
  publisher = {American Physical Society},
  doi = {10.1103/PhysRevB.52.R3876},
  url = {https://link.aps.org/doi/10.1103/PhysRevB.52.R3876}
}

@article{Takigawa2004,
author = {Takigawa ,Mitsuaki and Ichioka ,Masanori and Machida ,Kazushige},
title = {Quasiparticle {S}tructure in {A}ntiferromagnetism around the {V}ortex and {N}uclear {M}agnetic {R}elaxation {T}ime},
journal = {J. Phys. Soc. Jpn.},
volume = {73},
number = {2},
pages = {450-458},
year = {2004},
doi = {10.1143/JPSJ.73.450},
URL = {https://doi.org/10.1143/JPSJ.73.450}}

@article{Duneau1985,
  title = {Quasiperiodic Patterns},
  author = {Duneau, Michel and Katz, Andr\'e},
  journal = {Phys. Rev. Lett.},
  volume = {54},
  issue = {25},
  pages = {2688--2691},
  numpages = {0},
  year = {1985},
  month = {Jun},
  publisher = {American Physical Society},
  doi = {10.1103/PhysRevLett.54.2688},
  url = {https://link.aps.org/doi/10.1103/PhysRevLett.54.2688}
}

@article{Tomayeva2026,
  title = {Temperature-driven structural crossover of vortex matter in a {P}enrose quasicrystal},
  author = {Tomayeva, Munisa A. and Neverov, Vyacheslav D. and Shanenko, Arkady A. and Vagov, Alexei and Krasavin, Andrey V.},
  journal = {Phys. Rev. B},
  volume = {114},
  issue = {2},
  pages = {024516},
  numpages = {6},
  year = {2026},
  month = {Jul},
  publisher = {American Physical Society},
  doi = {10.1103/14xk-b4yd},
  url = {https://link.aps.org/doi/10.1103/14xk-b4yd}
}

@article{Junmo2025,
  title = {Interaction-tuned pattern-selective superconductivity: {A}pplication to the dodecagonal quasicrystal},
  author = {Jeon, Junmo and Lee, SungBin},
  journal = {Phys. Rev. B},
  volume = {112},
  issue = {17},
  pages = {174511},
  numpages = {7},
  year = {2025},
  month = {Nov},
  publisher = {American Physical Society},
  doi = {10.1103/6b35-y4zj},
  url = {https://link.aps.org/doi/10.1103/6b35-y4zj}
}

@article{Tada2025,
  title={Spin susceptibility in quasicrystal superconductor with spin-orbit interactions},
  author={Tada, Yasuhiro},
  journal={arXiv:2512.15093},
  year={2025},
  url={https://arxiv.org/abs/2512.15093v1}
}

\end{document}